\documentclass[epj]{svjour}
\usepackage{bbm,bm,graphicx,mathtools,color,slashed}
\usepackage{amssymb,amsmath}
\usepackage[caption=false]{subfig}
\usepackage[colorlinks=true, pdfstartview=FitV, linkcolor=red, citecolor=blue, urlcolor=blue]{hyperref}

\graphicspath{{./figures/}}

\newcommand{\bs}{\boldsymbol}

\begin{document}
\title{Chiral magnetic effect without chirality source in asymmetric Weyl semimetals}
\author{Dmitri E. Kharzeev\inst{1,2,3} \and Yuta Kikuchi\inst{2,4} \and Ren\'e Meyer\inst{5}% etc
}                     % Do not remove
\offprints{}          % Insert a name or remove this line
\institute{Department of Physics and Astronomy, Stony Brook University, Stony Brook, New York 11794-3800, USA \and 
Department of Physics, Brookhaven National Laboratory, Upton, New York 11973-5000 \and
RIKEN-BNL Research Center, Brookhaven National Laboratory, Upton, New York 11973-5000 \and
Department of Physics, Kyoto University, Kyoto 606-8502, Japan \and
Institute of Theoretical Physics and Astrophysics, University of W\"urzburg, 97074 W\"urzburg, Germany}
\date{Received: date / Revised version: date}
\abstract{
We describe a new type of the Chiral Magnetic Effect (CME) that should occur in Weyl semimetals with an asymmetry in the dispersion relations of the left- and right-handed chiral Weyl fermions. In such materials, time-dependent pumping of electrons from a non-chiral external source can generate a non-vanishing chiral chemical potential. This is due to the different capacities of the left- and right-handed (LH and RH) chiral Weyl cones arising from the difference in the density of states in the LH and RH cones. The chiral chemical potential then generates, via the chiral anomaly, a current along the direction of an applied  magnetic field even in the absence of an external electric field. The source of chirality imbalance in this new setup is thus due to the band structure of the system and the presence of (non-chiral) electron source, and not due to the parallel electric and magnetic fields. We illustrate the effect by an argument based on the effective field theory, and by the chiral kinetic theory calculation for a rotationally invariant Weyl semimetal with different Fermi velocities in the left and right chiral Weyl cones; we also consider the case of a Weyl semimetal with Weyl nodes at different energies. We argue that this effect is generically present in Weyl semimetals with different dispersion relations for LH and RH  chiral Weyl cones, such as ${\rm SrSi_2}$ recently predicted as a Weyl semimetal with broken inversion and mirror symmetries, as long as the chiral relaxation time is much longer than the transport scattering time.
\PACS{
      {PACS-key}{discribing text of that key}   \and
      {PACS-key}{discribing text of that key}
     } % end of PACS codes
} %end of abstract
\maketitle
\section{\label{sec:1}
Introduction}

The chiral magnetic effect (CME) \cite{Fukushima:2008xe} (see \cite{Kharzeev:2013ffa,Kharzeev:2012ph} for reviews and additional references) is a non-dissipative quantum transport phenomenon induced by the chiral anomaly in the presence of an external magnetic field. It has been predicted to occur in Dirac and Weyl semimetals (DSMs/WSMs) \cite{Fukushima:2008xe,Son:2012wh,Son:2012bg,Zyuzin:2012tv,Basar:2013iaa,vazifeh2013electromagnetic,goswami2013axionic}, and has been recently experimentally observed through the measurement of negative longitudinal magnetoresistance in DSMs \cite{Li:2014bha,kim2013dirac,xiong2015evidence,li2015giant} as well as WSMs  \cite{huang2015observation,wang2015helicity,zhang2015observation,yang2015observation,shekhar2015large,yang2015chiral}. In these materials, the electric current flows along the  external magnetic field $\bs{B}$ in the presence of a chiral chemical potential $\mu_5$, 
\begin{align}\label{eq1}
  \bs{j}_\mathrm{CME} = \frac{e^2}{2\pi^2}\ \mu_5 \bs{B}.
\end{align}
The chiral chemical potential can be formally introduced through the axion field linearly dependent on time, $\theta(t,\bs{x}) = - 2 \mu_5 t$; Eq.~(\ref{eq1}) then emerges \cite{Kharzeev2010} as a direct consequence of the chiral anomaly from the Chern-Simons  effective action that will be considered in Section {\ref{sec:2}}. 
The CME \eqref{eq1} relies on a source of chirality to generate the chiral chemical potential $\mu_5$. In the conventional realization of CME in DSMs and WSMs, the external electric and magnetic field provide such a source via the chiral anomaly,
\footnote{For the purpose of the conventional chiral anomaly induced chiral magnetic effect the two overlapping Weyl cones of a DSM contribute in the same way as for a WSM. We will see that this is not the case for the asymmetric CME discussed in this work, which vanishes in DSMs with exactly coinciding dispersion relations such as e.g. ZrTe${}_5$ \cite{Li:2014bha}.} 
\begin{align}\label{eqAnomaly}
  \frac{d\rho_5}{dt} = \frac{e^2}{4\pi^2\hbar^2 c}\bs{E}\cdot\bs{B} - \frac{\rho_5}{\tau_5},
\end{align}
where $\tau_5$ is a time scale describing chiral relaxation processes present due to spin impurities, edge modes (Fermi arcs) and band mixing (described in the effective field theory by higher-dimension ultraviolet (UV) operators). With the chiral relaxation term, \eqref{eqAnomaly} allows for a time independent solution with finite $\rho_5$ if and only if the electric field has a component parallel to the magnetic field. The induced chiral chemical potential then gives rise, via \eqref{eq1}, to the observed longitudinal magnetoconductivity growing as a square of magnetic field .

\begin{figure}
 \begin{center}
 \includegraphics[width=8.5cm]{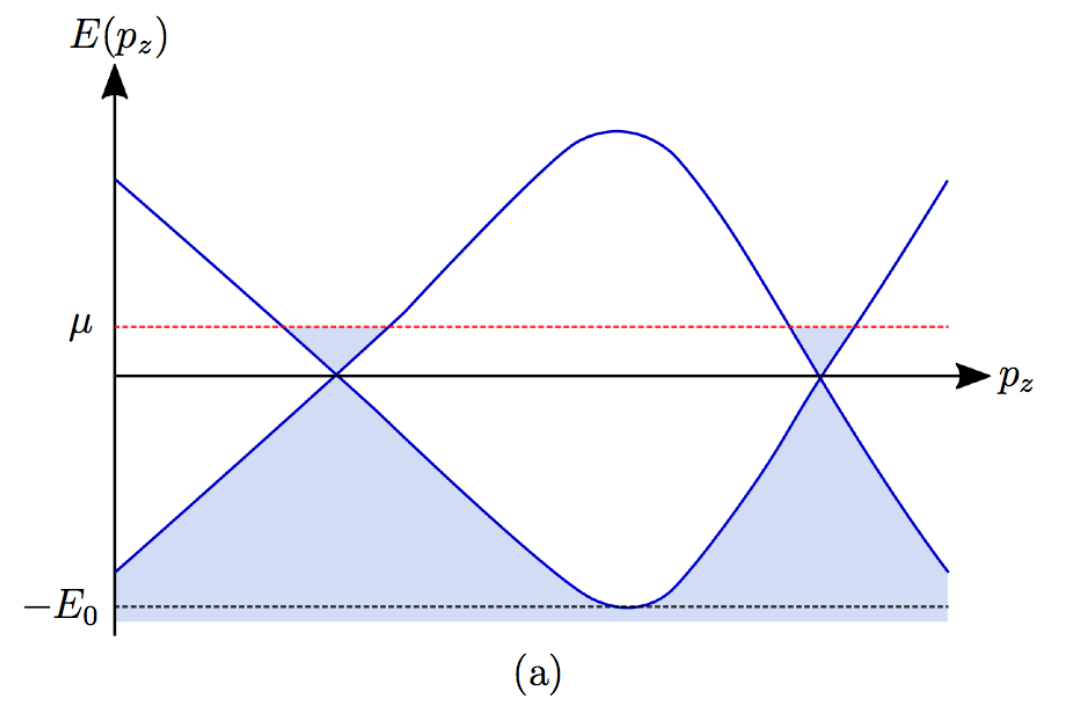}
 \includegraphics[width=8.5cm]{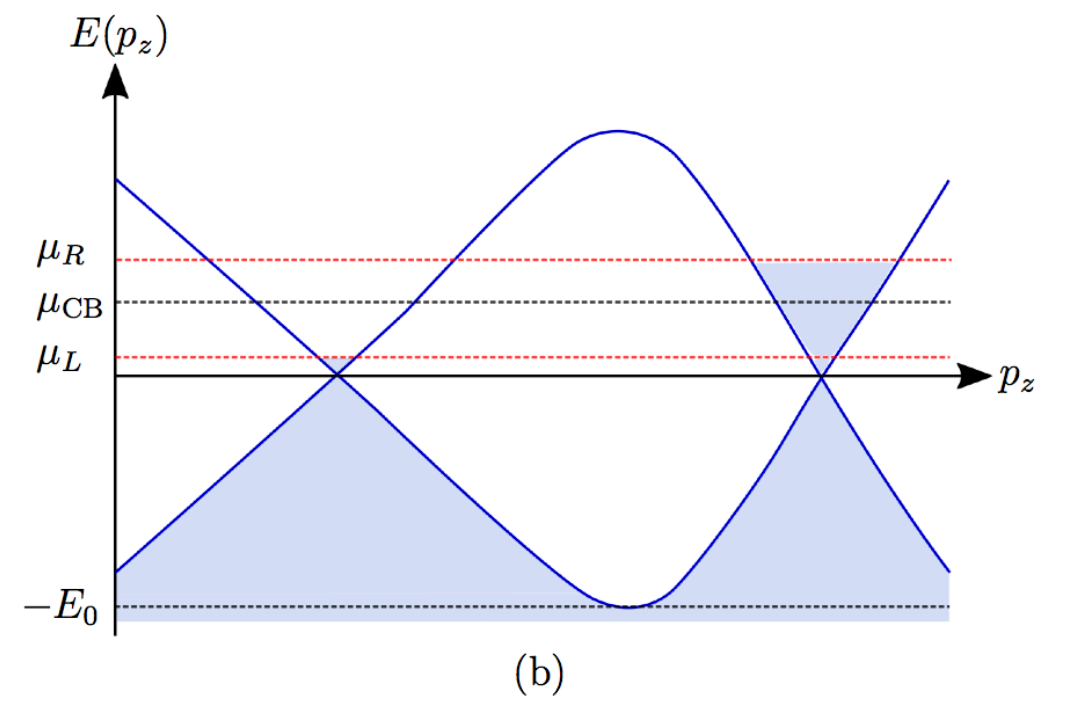}
 \caption{(Color online)
Schematic band structures of the asymmetric WSM possessing a pair of Weyl cones with different Fermi velocities for left and right. Fermi-sea contribution is regularized by a cutoff $-E_0$ determined by a bottom of the lower band. (a) Chirally balanced, i.e., equilibrium state with Fermi energy $E_F$. (b) Chirally imbalanced, i.e., non equilibrium state with different Fermi energies $E_{F,L}$ and $E_{F,R}$ for left and right Weyl nodes.}
 \label{fig:1}
 \end{center}
\end{figure}

\vskip0.3cm
Recently, a new class of Weyl semimetals with broken reflection and inversion symmetries was predicted, with 
${\rm SrSi_2}$ as an example \cite{huang2016new}. 
For brevity, we will refer to these materials as {\it asymmetric WSMs} (aWSMs). The authors of \cite{huang2016new} pointed out that aWSMs, in which the Weyl points with different chiral charges have different dispersion relations, are well suited for the studies of CME. 

In this paper we point out that the aWSMs indeed enable a new type of the CME  that does not require an electric field, but instead is driven by an external source of electrons (we will call it {\it asymmetric CME}, aCME). The induced aCME current is a direct signature of the chiral anomaly responsible for Eq. \eqref{eq1}. 

The chirality source is built into the band structure of the aWSM by means of asymmetric, i.e. non-identical, dispersion relations of the left and right chiral Weyl cones, see Fig.~\ref{fig:1} for illustration of the special case of rotationally invariant Weyl cones with different Fermi velocities. The asymmetry in the dispersion relations leads to a different density of states (per unit chemical potential) in the left and right Weyl cones, and hence to a different capacity for quasi-particles of different chiralities. 

The chirality imbalance can then be generated by pumping the system with a non-chiral time-dependent AC current, %(such as the one coming from a household electric plug), 
changing the chemical potential in  the left and right chiral Weyl cones at a different rate due to their different capacities, which in turn induces a chiral chemical potential leading to the aCME current \eqref{eq1}. 
Admittedly, the population of left and right Weyl cones will depend on detailed realization of the electron pumping. {We discuss two possible mechanisms below.}  Nevertheless, since there is no symmetry between the dispersion relations in the left and right cones, and the process is driving the system out of equilibrium, we expect that the chemical potentials in left and right Weyl cones will in general be different. {Once created, this population asymmetry will persist in a surface layer of thickness determined by the chirality 
flipping length ($v_F \tau_5$ in the ballistic regime, or $\sqrt{D \tau_5}$ in the diffusion regime). Experimentally, in CdAs$_3 $ \cite{Zhang:2017dd}
it is known to be $2$-$3 \mu$m.  The criterium of chirality equilibration is the system size being much larger than this length scale. We hence expect our effect to persist for films up to that thickness, which otherwise are known to  behave as bulk Weyl semimetals. The population asymmetry} will then amount to the chiral chemical potential driving the chiral magnetic current (\ref{eq1}). Let us illustrate this by assuming that the charge is pumped into the left and right cones with the same rate.
As an analogue of the effect, imagine two water buckets with different base areas. Obviously, the thinner bucket with smaller base area will fill up at a faster rate compared to the thicker one, if the same water inflow is applied to both. The difference in the water levels in ``left" and ``right" buckets in this analogy corresponds to the chiral chemical potential. 

If one describes the physics of electron pumping through the contacts by using the Fermi golden rule (i.e. assuming that the square of the amplitude of electron penetration into the material can be factorized from the density of final states), then the chiral chemical potential would vanish. However, the penetration amplitude will depend on spin if spin-polarized contacts are used - in this case, the chiral chemical potential will not vanish. The chirality dependence in the electron pumping amplitude can also arise from the Fermi arcs, that are chiral states localized on the surface. {In this case, the contact material should not break the inversion symmetry by itself, so that an equal number of left- and right-handed carriers 
is injected into the surface states. This can be realized with an ordinary metal with a large Fermi surface, i.e. the contact material does not need to admit chiral quasiparticles itself. The tunneling into the Fermi arc states will then generate the chirality imbalance.} The physics of the interface between aWSMs (and WSMs in general) and conventional metals still has to be understood -- however, in general we expect that the AC pumping of electrons into aWSMs can generate the chiral chemical potential in these systems. Lacking the theory of this interface, we cannot reliably predict the value of the chiral chemical potential, and thus the magnitude of the chiral magnetic current. However, the very existence of this current parallel to the magnetic field and orthogonal to the applied electric field, and the fact that (as will be shown below) it is shifted in phase by $\pi/2$ relative to the driving voltage, is a firm prediction of our approach that can be tested in experiment.

%Insisting on Lorentz invariance, possible representations of the Clifford algebra have been completely classified 
%in all dimensions. Three types of irreducible representations are known to exist: complex Dirac fermions, chiral Weyl fermions (in even dimensions), and real Majorana fermions. In condensed matter systems, all of these representations have been shown to emerge in the electronic band structure of solids. 
%Weyl semimetals (WSMs) are a realization of massless Weyl fermions as low-energy excitations of electrons, which emerge at distinct points in the Brillouin zone where conduction and valence bands touch with the linear dispersion of a left or right chiral Weyl fermion. Dirac semimetals (DSMs) are WSMs in which the left and right chiral band touchings happen at the same point in the Brillouin zone with (up to chirality) identical dispersion relations. In this case, the left and right chiralities pair up to form a 3+1-dimensional Dirac spinor representation. These materials recently have attracted much attention as topological semimetals, in particular due to their interesting anomaly induced non dissipative transport phenomena such as the CME \eqref{eq1}.

\begin{figure}
 \begin{center}
 \includegraphics[width=8.5cm]{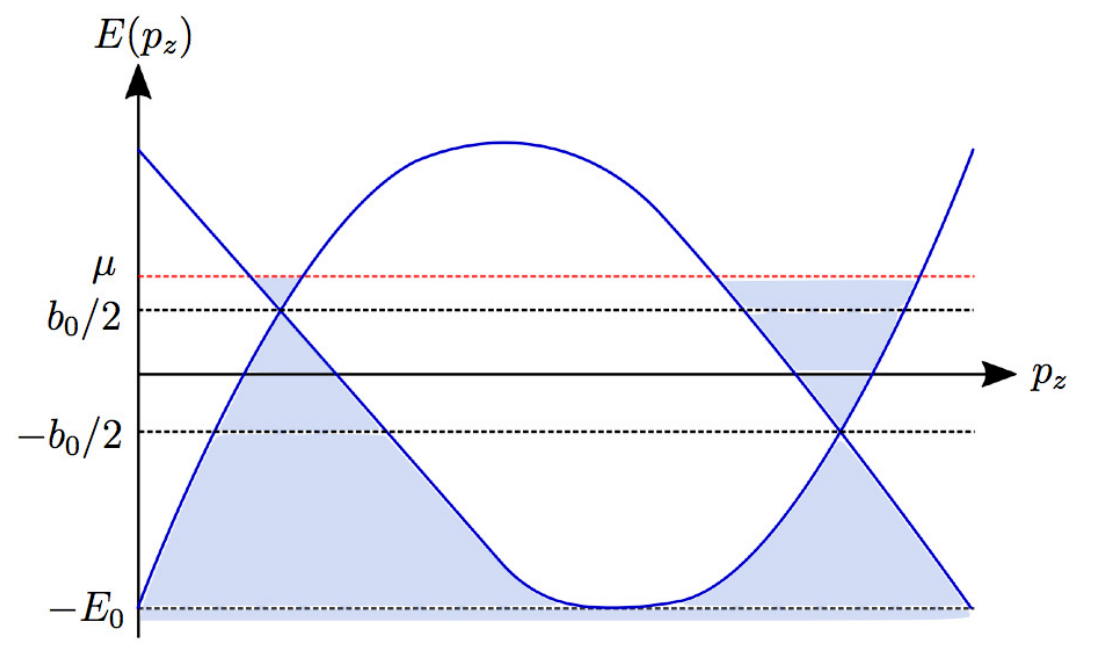}
 \caption{(Color online)
Schematic band structures of the asymmetric WSM, which possesses a pair of Weyl cones with finite energy difference $b_0$}
 \label{fig:4}
 \end{center}
\end{figure}

Weyl and Dirac semimetals can be classified in terms of the discrete symmetries of the Lorentz group \cite{murakami2007phase,Kharzeev:2011ds,yang2014classification,burkov2015chiral}: parity (P), that coincides with inversion (I) in odd number of spatial dimensions, and time reversal (T). Separating left and right chiral Weyl nodes along a vector $\bs{b}$ in momentum space breaks T but preserves P, while separating the nodes in energy only (but not in momentum) by an amount $b_0$ preserves T and breaks P. A WSM hence necessarily breaks either P or T, or both, as in Fig.~\ref{fig:4}. A DSM, on the other hand,  preserves both P and T. Left- and right-handed Weyl fermions are bound to exist in pairs in a lattice system due to the Nielsen-Ninomiya theorem \cite{nielsen1981no,nielsen1981absence,nielsen1981absence2} or, in other words, since the Berry curvature of the Brillouin zone as a whole has to vanish \cite{Kiritsis:1986re}. 

 The above classification in terms of P and T assumes identical dispersion relations for the left- and right-handed chiral particles.  Parity exchanges left and right chiralities, and hence is broken explicitly if the dispersion relations of the left and right chiral Weyl fermions are not identical \footnote{More precisely, since parity (P) in 3+1 dimensions acts on the spatial coordinates and momenta as $x^i \mapsto - x^i$ and $k^i\mapsto - k^i$ ($i=1,2,3$), respectively, the condition for parity invariance for the low energy dispersion relation is $\omega_L(k_i) = \omega_R(-k_i)$, i.e. for a left chiral Weyl node at spatial momentum $\bs{k}$, there exists an identical right chiral node at momentum $-\bs{k}$. Time reversal (T) on the other hand sends $\bs{k}\mapsto - \bs{k}$ but does not exchange chiralities, and hence Weyl nodes must exist in pairs of the same chirality but at opposite momenta in the Brillouin zone.}. 

Asymmetric Weyl semimetals (aWSMs) possess such a parity non-invariant pair of  Weyl cones with opposite chiralities. A specific realization of aWSM has been proposed recently in \cite{huang2016new}.
 aWSMs constitute a new distinct class of WSMs, and topological semimetals in general. %\footnote{Note that even a Dirac semimetal can be asymmetric if both P and T are broken. In this case, the  left and right Weyl nodes sit at the origin of the Brillouin zone, but the dispersion relations are not the same, breaking Kramers degeneracy and hence T, as well as P. While in ordinary WSMs the large chiral relaxation time $\tau_5$ is related to the distance of the Weyl nodes in either momentum or energy, we expect aWSMs to have a suppressed chiral relaxation rate also due to the asymmetry between the Weyl nodes.}. 
 Since in  condensed matter systems Lorentz invariance and in particular rotational invariance is not a fundamental  but rather an emergent symmetry at low energies, aWSMs can be expected to occur quite generically. {Our new effect requires the absence of inversion symmetry as well as of any mirror planes, as reflections invert the chiralities of the Weyl nodes and hence of any asymmetry in their dispersion relations. We discuss the crystal symmetries fulfilling these requirements in sec.~\ref{sec:6}.}

An interesting (and well controlled theoretically) way of breaking parity and Lorentz symmetry \footnote{Here, Lorentz symmetry is preserved for each Weyl cone separately, but broken since the left and right handed Weyl fermions do not have the same velocities, i.e. do not travel on an overall universal light cone.} while preserving rotational invariance is to introduce different Fermi velocities for the left and right Weyl cones, as depicted schematically in Fig.~\ref{fig:1}.  In high energy physics, this would correspond to a (rather pathological) universe in which left- and right-handed massless chiral electrons move at a different speed, so there is no unique speed of light.

From the point of view of chiral kinetic theory it is clear that an asymmetric deformation of the Weyl cones does not affect anomalous transport phenomena such as the CME \eqref{eq1} since the singular structure of the band touchings in momentum space, the Berry monopole, is invariant under such small deformations. We will indeed confirm in Sec.~\ref{sec:3} and App.~\ref{app:4} that in the cases considered here the chiral kinetic expression for the CME current remains of the form \eqref{eq1}. 
It seems however nontrivial in terms of field theory, where the anomaly is caused by subtle ultraviolet (UV) properties, and the absence of Lorentz invariance in the UV a priori may lead to a modification of the anomaly. For this reason, we calculate in Sec.~\ref{sec:2} the effective action in the case of different Fermi velocities for left and right chiral Weyl nodes, and show that \eqref{eq1} is unchanged also in this approach. Although this does not constitute a proof for the general case, it gives us confidence that, as expected from the anomaly matching arguments \cite{hooft1980naturalness,Frishman:1980dq}, the chiral magnetic current should always be of form \eqref{eq1} as long as  Lorentz symmetry is restored for each Weyl node separately at low energies.

In Sec.~\ref{sec:4} we discuss the proposed mechanism of the generation of the chiral chemical potential in aWSMs.
 In Sec.~\ref{sec:5} (c.f. Fig.~\ref{fig:exp}), we describe a possible experimental realization of the proposed mechanism. First, by applying an AC voltage to the grounded sample, we induce a time-dependent change in the chemical potential $\mu(t)$. The different capacities of the left and right handed Weyl nodes of an asymmetric WSM, described by the different equations of state $\rho_{L/R}(\mu_L/R,T,\dots)$ \footnote{The dots represent additional parameters such as the different Fermi velocities for each chirality, or the different tilting parameters for type I/II Weyl semimetals. They also include necessary physical cutoffs, as well as the Dirac sea cutoff.}, do allow for a non vanishing chiral charge density $\rho_5 = \rho_L-\rho_R$ even for vanishing chiral chemical potential $\mu_5=(\mu_L-\mu_R)/2=0$. The contributions from $\rho_{L/R}$ to $\rho_5$ in this case do not exactly cancel each other due to the spectral asymmetry between the left and right Weyl nodes, as e.g. for unequal Fermi velocities depicted in Fig.~\ref{fig:1},
\begin{align}\label{eq2}
 \rho_{L/R}  =\frac{\mu_{L/R}^3+E_0^3}{6\pi^2v_{L/R}^3}.
\end{align}
This is also true in equilibrium (i.e in a time independent configuration with $\mu_5=0$), in which an asymmetric WSM has a non-vanishing chiral charge density $\rho_5 \equiv \rho_L - \rho_R$ that does not vanish even at $\mu_5=0$, c.f.~(\ref{eq2}) for the case of $v_L \neq v_R$. In equilibrium and in the absence of the chiral anomaly, however, this chiral charge density is conserved and does not lead to any electrical transport. 
\vskip0.3cm

This paper is organized as follows.
In Sec.~\ref{sec:2},
the effective action describing anomalous electric transport phenomena such as the CME are derived in the case of an asymmetric WSM induced by different Fermi velocities for the left and right chiral Weyl nodes. 
In Sec.~\ref{sec:3},
we derive the aCME current at zero temperature from chiral kinetic theory and show that it reproduces the result of the field theoretic derivation of Sec.~\ref{sec:2}.
Sections~\ref{sec:4} and \ref{sec:5} are the main part of this paper. 
In Sec.~\ref{sec:4}, we %specify the constitutive relation between the chiral chemical potential and chiral density, which is peculiar to the asymmetric WSM due to the different electron capacities between two Weyl nodes. Based on the relation, we point out a novel way of generating the chirality imbalance and the corresponding CME current. 
derive from the equation of state \eqref{eq2} for unequal Fermi velocities the relation between applied pumping rate and induced chiral chemical potential which in turn induces the aCME current. 
In Sec.~\ref{sec:5}, we discuss an experimental setup needed to observe the aCME in asymmetric WSMs 
and give a quantitative estimate of the current density. 
Sec.~\ref{sec:6} is devoted 
to a summary and conclusions. In particular we discuss a class of asymmetric WSMs (we call them chirally imbalanced materials) that should exhibit the aCME.  
\vskip0.3cm

Technical details are relegated to several Appendices. 
In App.~\ref{app:1}, 
we present the detailed calculation of the index necessary to arrive at the effective action of Sec.~\ref{sec:2}. 
In App.~\ref{app:2}, we extend our chiral kinetic theory calculations of the asymmetric CME current discussed in Sec.~\ref{sec:3} and \ref{sec:4} to finite temperatures.  We find that for the case of asymmetric Fermi velocities the finite temperature contributions cancel in the final result, and comment on the non generality of such cancellations.  In App.~\ref{app:4} we derive and estimate the asymmetric CME in WSMs with Weyl nodes separated not only in momentum but also in energy, as depicted in Fig.~\ref{fig:4}. 
The conversion from natural into Gaussian units and the estimate of the asymmetric CME currents are summarized in App.~\ref{app:3}. 

\medskip

%%%%%%%%%%%%%%%%%%%%%%%%%%%%%%%%%%%%%%%%%%%%%%%%%%%%

\section{\label{sec:2}
Chern-Simons effective action and topological response}

WSMs are topological materials possessing chiral Weyl nodes separated in momentum and/or energy; the anomalous response of these materials to external field is described by the Chern-Simons action.
As already discussed in Sec.~\ref{sec:1}, the CME \eqref{eq1} is induced by the chiral chemical potential $\mu_5$. As will be reiterated below, the chiral anomaly indeed generates the Chern-Simons term which leads to the CME current \eqref{eq1}.
Conventional WSMs (i.e. the ones possessing identical dispersion relations for left- and right-handed fermions) are described by the following action for a four component Dirac spinor $\psi$ in Euclidean signature
\begin{align}
 \label{eq:action}
 &S = \int d\tau d\bs{r} \bar{\psi}i (\slashed{D} + i\slashed{b}\gamma_5)\psi
 \nonumber\\
 &=\int d\tau d\bs{r}
 \bar{\psi}\!
  \begin{pmatrix}
   0 &  i\bar{\sigma}^\mu (D_\mu \!- \! ib_\mu)\!-\!\mu_5
   \\
   i\sigma^\mu (D_\mu \!+ \! ib_\mu)\!+\!\mu_5 & 0
  \end{pmatrix}
 \!\psi,
\end{align}
where $\slashed{D}\equiv\gamma^\mu D_\mu$ and $\slashed{b}\equiv\gamma^\mu b_\mu$. $\gamma^\mu$ $(\mu=0,1,2,3)$ denotes Dirac matrices, and $\sigma^\mu=(1,\sigma^i)$ and $\bar{\sigma}^\mu=(1,-\sigma^i)$ are defined via the Pauli matrices $\sigma^i$ $(i=1,2,3)$. 
Here, $b^\mu$ represents the separation between the two Weyl nodes in energy-momentum space. It should be emphasized \cite{Basar:2013iaa} that $\mu_5$, which is induced by non equilibrium processes, contributes to the CME response \eqref{eq1}, while $b_0$ is a part of parameterizations of the band structure and hence does not directly lead to persistent current via \eqref{eq1} \footnote{As shown in App.~\ref{app:4}, $b_0$ will however be responsible for an aCME response in the presence of an external non chiral pumping current.}. 
By performing the chiral gauge transformations 
\begin{align}
 \label{eq:chiral_trans_1}
 &\psi_{L/R} \rightarrow e^{-i\theta(x)\gamma_5/2}\psi_{L/R}=e^{-i\theta(x)(\gamma_5\pm1)/4}\psi_{L/R},
 \\
 \label{eq:chiral_trans_2}
 &\bar{\psi}_{L/R} \rightarrow \bar{\psi}_{L/R}e^{-i\theta(x)\gamma_5/2}=\bar{\psi}_{L/R}e^{-i\theta(x)(\gamma_5\mp1)/4},
\end{align}
with $\theta=-2\mu_5t$ in the action \eqref{eq:action}, $\mu_5$ terms are rotated away and the Chern-Simons action\begin{align}
 \label{eq:CSaction} 
 S_{\mathrm{CS}}&=\frac{e^2}{32\pi^2}\int d^4x\theta(x)\epsilon^{\mu\nu\alpha\beta}F_{\mu\nu}F_{\alpha\beta}
\end{align}
 is generated. 

Following the discussion given in \cite{fujikawa2004path,Zyuzin:2012tv}, we now consider an effective action for the asymmetric WSM where the Fermi velocities of left and right chiral fermions respectively, $v_L$ and $v_R$, take different values (c.f. Fig.~\ref{fig:1}). In this case, the action \eqref{eq:action} is modified as 
\begin{align}
 \label{eq:action1}
 &S =\int d\tau d\bs{r} \bar{\psi}
  \begin{pmatrix}
   0 & i\tilde{D}_{R}+\mu_5
   \\
   i\tilde{D}_{L}-\mu_5 & 0
  \end{pmatrix}
 \psi.
\end{align}
Here, we have defined covariant derivatives
\begin{align}
 &\tilde{D}_{L}= (D_0 + ib_0)-v_L\sigma^i (D_i + ib_i),
 \nonumber\\
 &\tilde{D}_{R}= (D_0 - ib_0)+v_R\bar{\sigma}^i (D_i - ib_i),
\end{align}
which are 2$\times$2 component matrices.
Eq.~\eqref{eq:action1} can be rewritten with use of four-component Weyl spinors $\psi_L$ and $\psi_R$ satisfying $\gamma_5\psi_{L/R}=\pm\psi_{L/R}$,
\begin{align}
 \label{eq:action2}
 S = \int d\tau d\bs{r} \big[&\bar{\psi}_L i(\slashed{D}_L-i\mu_5\gamma^0\gamma_5)\psi_L
 \nonumber\\
 +&\bar{\psi}_R i(\slashed{D}_R-i\mu_5\gamma^0\gamma_5)\psi_R\big],
\end{align}
where covariant derivatives are now defined as 
\footnote{With this definition of the 4$\times$4 Dirac operator it is known that the following procedure yields the covariant anomaly in the system where the chiral Weyl fermions couple to the external gauge field. See \cite{fujikawa2004path} for detailed explanation. }
\begin{align}
 \label{eq:DL_DR}
 \slashed{D}_{L/R}= \big[\gamma^0(D_0 + ib_0\gamma_5)+v_{L/R} \gamma^i (D_i + ib_i\gamma_5)\big].
\end{align}
We will now show that the chiral gauge transformation \eqref{eq:chiral_trans_1} and \eqref{eq:chiral_trans_2} 
yields the conventional four dimensional Chern-Simons term \eqref{eq:CSaction} despite the asymmetric deformation of the Dirac action.
In particular, the chiral chemical potential term in \eqref{eq:action2} can be rotated away by a chiral rotation with $\theta=-2\mu_5t$, in which case \eqref{eq:action2} describes the magnetic response of the aWSM to non vanishing $\mu_5$.

Let us first focus on the left-handed fermion part of the action \eqref{eq:action2}.
$\psi_L$ and $\bar{\psi}_L$ can be decomposed in terms of Dirac eigenfunctions $\{\phi_n\}$, such that $\slashed{D}_L\phi_n=\lambda_n\phi_n$, as
\begin{align}
 \psi_L=\sum_n \phi_n c_n, \ \ \ 
 \bar{\psi}_L=\sum_n \phi^*_n \bar{c}_n.
\end{align}
%It is noted that $\{\phi_n\}$ in the expansion of $\psi_L$ includes right-chiral modes because of $[\slashed{D}_L,\gamma_5]\ne0$.
Infinitesimal chiral rotations of $c$ and $\bar{c}$ are specified from Eqs.~\eqref{eq:chiral_trans_1} and \eqref{eq:chiral_trans_2},
\begin{align}
 c'_n&=\sum_m U_{nm}c_m, 
 \nonumber\\
 &U_{nm}=\delta_{nm}-\frac{i}{4}\int d^4x\phi^*_n(x)\theta(x)(\gamma_5\pm1)\phi_m(x),
 \\
 \bar{c}'_n&=\sum_m V_{mn}\bar{c}_m,
 \nonumber\\
 &V_{mn}=\delta_{mn}-\frac{i}{4}\int d^4x\phi^*_m(x)\theta(x)(\gamma_5\mp1)\phi_n(x).
\end{align}
Henceforth, the Jacobian of the chiral gauge transformation becomes
\begin{align}
 \label{eq:determinant}
 \det(U^{-1}V^{-1})=\exp\left(\frac{i}{2}\int d^4x\theta(x)\sum_n\phi^*_n(x)\gamma_5\phi_n(x)\right).
\end{align}
The right hand side of the above equation includes the expression of the index for a single left chiral fermion, \begin{align}
 \label{eq:left_CS}
 \sum_n\phi^*_n(x)\gamma_5\phi_n(x) = \frac{e^2}{32\pi^2}\epsilon^{\mu\nu\alpha\beta}F_{\mu\nu}F_{\alpha\beta}.
\end{align}
where $\epsilon^{\mu\nu\alpha\beta}$ is the antisymmetric tensor and the field strength is defined by $F_{\mu\nu}=\partial_\mu A_\nu-\partial_\nu A_\mu$. The detailed calculation of the index is presented in App.~\ref{app:1}.
Notice that \eqref{eq:left_CS} does not depend on the Fermi velocity $v_L$, which cancels out during the calculation (App.~\ref{app:1}).  The chiral gauge transformation for the right chiral action gives the same expression as \eqref{eq:determinant} and \eqref{eq:left_CS}. Therefore, by combining the Jacobians caused by the left and right chiral gauge transformations we obtain the usual Chern-Simons term as an effective action \eqref{eq:CSaction}
\begin{align}
 \label{eq:CSaction2} 
 S_{\mathrm{CS}}&=-\frac{e^2}{8\pi^2}\int d^4x\epsilon^{\mu\nu\alpha\beta}\partial_\mu\theta(x)A_{\nu}\partial_\alpha A_{\beta},
\end{align}
where we have integrated by parts in \eqref{eq:CSaction} and neglected the surface term.
By taking the derivative with respect to $A_i$ and setting $\theta=-\mu_5t$, we obtain the the anomaly induced current
\begin{align}
 \label{eq:anom_current}
 j_i &=\frac{\delta S_\mathrm{CS}}{\delta A_i}= \frac{e^2}{2\pi^2}\mu_5\epsilon^{0i\alpha\beta}\partial_\alpha A_\beta
 =\frac{e^2}{2\pi^2}\mu_5B_i.
\end{align}
This expression for the chiral magnetic current agrees with that of conventional WSMs. We will see in the next section that chiral kinetic theory
%, which is valid in the semiclassical regime of small magnetic field $|B|/E_{F,{L/R}} \ll 1$, 
yields the same result.

%%%%%%%%%%%%%%%%%%%%%%%%%%%%%%%%%%%%%%%%%%%%%%%%%%%%

%%%%%%%%%%%%%%%%%%%%%%%%%%%%%%%%%%%%%%%%%%%%%%%%%%%%

\section{\label{sec:3} 
Chiral Kinetic Approach}

In this section, we use chiral kinetic theory to describe the CME in asymmetric WSMs with different Fermi velocities. 
Chiral kinetic theory allows to describe the non equilibrium states considered here by incorporating the chiral anomaly via to geometric Berry phase (Berry curvature)~\cite{Son:2012wh,Stephanov:2012ki} of the Weyl nodes into kinetic theory.

In aWSMs, the excitations close to left and right chiral Weyl nodes are described by the following Hamiltonian,
\begin{align}
 H_{L/R} = \pm v_{L/R}\bs{\sigma}\cdot(\bs{p}\pm\bs{b}/2),
\end{align}
where $\bs{b}$ represents the spatial separation between the left and right Weyl nodes in momentum space. In this and following sections, we consider the effect of the different Fermi velocities $v_L$ and $v_R$, and the energy difference between a pair of Weyl nodes is set to zero ($b_0=0$) for simplicity. Although nonvanishing $b_0$ does not affect the result \eqref{eq:CME}, it plays an important role in inducing chiral chemical potential as briefly discussed in Sec.s~\ref{sec:4} and \ref{sec:5}; a more detailed discussion is presented in App.~\ref{app:4}.

Let us concentrate on the left-handed chiral part. 
Diagonalizing $H_L$ yields two eigenenergies, $\epsilon_\pm =\pm v_{L}|\bs{p}+\bs{b}/2|$ with corresponding eigenvectors
\begin{align}
 u_{\pm,\bs{p}}=&\frac{1}{\sqrt{2|\bs{p}+\bs{b}/2|(|\bs{p}+\bs{b}|\mp (p_z+b_z/2))}}
 \nonumber\\
 &\times\begin{pmatrix}
  (p_x+b_x/2)-\mathrm{i}(p_y+b_y/2) \\
  \pm|\bs{p}+\bs{b}/2|-(p_z+b_z/2)
 \end{pmatrix},
\end{align}
respectively.
The Berry connection $\bs{\mathcal{A}}$ and Berry curvatures $\bs{\Omega}$ for these states are given by
\begin{align}
 \bs{\mathcal{A}} &\equiv -\mathrm{i} u^\dagger_{\pm,\bs{p}} \nabla_{\bs{p}}u_{\pm,\bs{p}},
 \\
 \label{eq:Berry_curvature}
 \bs{\Omega} &\equiv \bs{\nabla}\times\bs{\mathcal{A}}
 =\pm \frac{\bs{p}+\bs{b}/2}{2|\bs{p}+\bs{b}/2|^3}.
\end{align}
Eq.~\eqref{eq:Berry_curvature} describes a geometric structure in momentum space. 
One can see that a monopole (antimonopole) is located at $\bs{p}=-\bs{b}/2$ corresponding to the positive (negative) energy part of the left Weyl cone. Similarly, the positive (negative) energy right Weyl cone gives an antimonopole (monopole) at $\bs{p}=\bs{b}/2$. The monopole charges due to the left and right Weyl nodes sum up to zero in accord with Nielsen-Ninomiya theorem \cite{nielsen1981no,nielsen1981absence,nielsen1981absence2}.

Notice that the chiral kinetic approach breaks down around the monopole singularities located at the nodal points of the Weyl cones where the quantum anomaly violates the semiclassical description.
To avoid these singularities, the typical momentum of the quasi particles described in kinetic theory should be outside the quantum region around the nodal points. For example, in near equilibrium systems at zero temperature, the momentum of the quasi particle is characterized by the chemical potential $\mu$, while the typical radius of quantum region is $\sqrt{|\bs{B}|}$. Therefore, $\sqrt{|\bs{B}|}\ll\mu$ should be satisfied, which implies that strong magnetic field invalidates the chiral kinetic approach.

In chiral kinetic theory, the CME current is expressed as \cite{Son:2012wh,Stephanov:2012ki,Chen:2012ca}
\begin{align}
 \label{eq:CMEcurrent}
 \bs{j}_\mathrm{CME} = e^2\bs{B}\int_{\bs{p}} f_{\bs{p}} \left(\bs{\Omega}\cdot\frac{\partial \epsilon_{\bs{p}}}{\partial\bs{p}}\right),
\end{align}
where $\int_{\bs{p}}\equiv\int \mathrm{d}^3\bs{p}/(2\pi)^3$ and $f_{\bs{p}}(x)$ is the one body distribution function.
Now, we calculate the left chiral contribution to the CME current at vanishing temperature. An extension to finite-temperature is straightforward (c.f. App.~\ref{app:2}).
The equilibrium distribution function at zero temperature is given by $\Theta(\mu_L-\epsilon_{\bs{p}})$. Eq.~\eqref{eq:CMEcurrent} is then evaluated as
\begin{align}
 \label{eq:CMEcurrent2}
 \bs{j}_{\mathrm{CME},L} &= e^2\bs{B}\int_{\bs{p}} \Theta(\mu_L-\epsilon_{\bs{p}}) \left(\bs{\Omega}\cdot\frac{\partial \epsilon_{\bs{p}}}{\partial\bs{p}}\right)
 \nonumber\\
 &= \frac{e^2\bs{B}}{4\pi^2}(\mu_L+E_0),
\end{align}
where $-E_0$ is the physical cutoff determined by the band structure as depicted in Fig.~\ref{fig:1} \cite{Basar:2013iaa}.
Similarly, the right chiral Weyl node yields 
\begin{align}
 \bs{j}_{\mathrm{CME},R} = -\frac{e^2\bs{B}}{4\pi^2}(\mu_R+E_0),
\end{align}
which has the opposite sign compared with the left contribution due to the opposite monopole charge. Therefore, the total CME current is given by
\begin{align}
 \label{eq:CME}
 \bs{j}_\mathrm{CME} = \frac{e^2\mu_5}{2\pi^2}\bs{B},
\end{align}
with the chiral chemical potential $\mu_5\equiv(\mu_L-\mu_R)/2$.
This expression agrees with that obtained from the effective CS action \eqref{eq:anom_current}.

%%%%%%%%%%%%%%%%%%%%%%%%%%%%%%%%%%%%%%%%%%%%%%%%%%%%

%%%%%%%%%%%%%%%%%%%%%%%%%%%%%%%%%%%%%%%%%%%%%%%%%%%%
\section{\label{sec:4} 
Pumping-induced chirality imbalance in asymmetric WSM}

In this section, we describe a way to induce the chirality imbalance by pumping electrons into the asymmetric WSM.
Let us assume for simplicity that  the left and right chiral charge densities increase at the same rate 
 (we will comment on the electron pumping mechanism in the next section).
 A detailed experimental setup will be discussed in Sec.~\ref{sec:5}. 
We assume that due to pumping the total particle number density is time-dependent, $\rho=\rho(t)$.
The particle number densities and chemical potentials for each Weyl node are related by 
\begin{align}
 \rho_{L/R} =  \int_{\bs{p}} \Theta(\mu_{L/R}-v_{F,L/R}|\bs{p}|)
 =\frac{\mu_{L/R}^3+E_0^3}{6\pi^2v_{L/R}^3},
\end{align}
and the chiral charge density is given by $\rho_5=\rho_L-\rho_R$.
In a {\it chirally balanced} state, defined by a state satisfying $\mu_L=\mu_R\equiv\mu_\mathrm{CB}(t)$, the left and right chiral particle densities are $\rho_{L/R}^\mathrm{CB}=\rho_{L/R}(\mu_{L/R}=\mu_\mathrm{CB})$ and the total number density $\rho=\rho_{L}^\mathrm{CB}+\rho_{R}^\mathrm{CB}$ is given by
\begin{align}
 \rho = \frac{\mu_\mathrm{CB}^3+E_0^3}{6\pi^2}\left(\frac{1}{v_{L}^3}+\frac{1}{v_{R}^3}\right)
 =\rho(t),
\end{align}
which leads to the following expression for the chiral density in the chirally balanced state,
\begin{align}
 \label{eq:rho5CB}
 \rho_5^\mathrm{CB} = -\rho(t)\frac{v_L^3-v_R^3}{v_L^3+v_R^3}+const.
\end{align}
Here we neglect a time independent constant involving the Dirac sea cutoff $E_0$. It is noteworthy that $\mu_5=\mu_L-\mu_R=0$ in the chirally balanced state, although $\rho_5^\mathrm{CB}$ is finite due to the different capacities between two Weyl cones.
Since the pumping current is not chiral, the chiral charge density is conserved,
\begin{align}
 \label{eq:pump5}
 \frac{d\rho_5}{dt}= 0.
\end{align} 
The anomalous contribution is absent because $\bs{E}\cdot\bs{B}=0$.
The chiral density $\rho_5$ can be decomposed as $\rho_5=\rho_5^\mathrm{CB}+\delta\rho_5$, respectively.
With the use of \eqref{eq:rho5CB}, \eqref{eq:pump5} can then be rewritten as 
\begin{align}
 \label{eq:deltarho5}
 \frac{d\delta\rho_5}{dt}&= \frac{d\rho(t)}{dt}\frac{v_L^3-v_R^3}{v_L^3+v_R^3} - \frac{\delta\rho_5}{\tau_5},
\end{align}
where we have added the last term on the right hand side to describe the chirality relaxation towards the chirally balanced state with chiral density $\rho_5^\mathrm{CB}$.

Competition between the generation of chirality imbalance induced by the particle pumping and the chiral relaxation results in a late time solution of \eqref{eq:deltarho5} at $t\gg \tau_5$. Aiming at describing an  experimental measurement, we assume oscillatory particle pumping, i.e., $\rho(t)=\bar{\rho}+\delta\rho(t)$ with $\delta\rho(t)\propto e^{i\omega t}$. 
Then, the late time solution of \eqref{eq:deltarho5} is
\begin{align}
\label{eq:deltarho5_2}
 \delta\rho_5 &= \frac{\tau_5}{1+i\omega\tau_5}\frac{d\rho(t)}{dt}\frac{v_L^3-v_R^3}{v_L^3+v_R^3},
\end{align}
We now relate this solution to the chiral chemical potential. 
To this end, we consider the quantity $v_L^3\delta\rho_L-v_R^3\delta\rho_R$, which can be converted in two different ways, 
\begin{align}
 \label{eq:rhomu1}
 &v_L^3\delta\rho_L-v_R^3\delta\rho_R
 =\frac{v_L^3-v_R^3}{2}\delta\rho+\frac{v_L^3+v_R^3}{2}\delta\rho_5
 \nonumber\\
 &=\frac{\tau_5}{1+i\omega\tau_5}\frac{d\rho(t)}{dt}\frac{v_L^3-v_R^3}{2},
 \\
 \label{eq:rhomu2}
 &v_L^3\delta\rho_L-v_R^3\delta\rho_R
 =\frac{\mu_{L}^3-\mu_{R}^3}{6\pi^2}
 =\frac{3\mu^2\mu_5-\mu_5^3}{3\pi^2}
 \nonumber\\
 &\simeq \frac{\mu^2\mu_5}{\pi^2} (1+ O((\mu_5/\mu)^2)).
\end{align}
where $\mu=(\mu_L+\mu_R)/2$ and $\mu_5=(\mu_L-\mu_R)/2$, and we have assumed that $\mu_5 \ll \mu$.
Combining these two equations yields the expression for the chiral chemical potential,
\begin{align}
 \label{eq:mu5i}
 \mu_5\simeq \frac{\pi^2(v_L^3-v_R^3)}{2\mu^2}\frac{\tau_5}{1+i\omega\tau_5}\frac{d\rho(t)}{dt}.
\end{align}
In addition, the relation between the particle number density and chemical potential becomes
\begin{align}
\label{eq:rho_mu}
&\rho(t) 
= \frac{1}{6\pi^2}\left(\frac{\mu_L^3+E_0^3}{v_{L}^3}+\frac{\mu_R^3+E_0^3}{v_{R}^3}\right)
\nonumber\\
&= \frac{\mu^3+E_0^3}{6\pi^2}\left(\frac{1}{v_{L}^3}+\frac{1}{v_{R}^3}\right)
\left(1+O\left(\frac{\Delta v}{V}\frac{\mu_5}{\mu}\right)\right).
\end{align}
where we have defined $\Delta v\equiv v_L-v_R$ and $V\equiv (v_L+v_R)/2$ and picked up the leading order of $\Delta v/V$ in the final expression.
Therefore, the chiral chemical potential is given by
\begin{align}\label{eq:mu5ii}
 \mu_5= \frac{(v_L^6-v_R^6)}{4v_L^3v_R^3}\frac{\tau_5}{1+i\omega\tau_5}\frac{d\mu}{dt}
 \simeq \frac{3}{2}\frac{\Delta v}{V}\frac{\tau_5}{1+i\omega\tau_5}\frac{d\mu}{dt}.
\end{align}
Let us reiterate that the chiral chemical potential is induced by the different Fermi velocities for left and right Weyl nodes in this mechanism, due to the different capacity of electron states in the two Weyl cones. The mechanism is distinct from the conventional one in the symmetric WSM, where the chiral chemical potential is generated by an external electric field component parallel to the external magnetic field, i.e. by $\bs{E}\cdot\bs{B}$.

Finally, inserting \eqref{eq:mu5ii} into \eqref{eq:CME}, we obtain the expression for the aCME current,
\begin{align}
 \label{eq:CME3}
 \bs{j}_\mathrm{CME} 
 &= \frac{3e^2}{4\pi^2}\frac{\Delta v}{V}\frac{\tau_5}{1+i\omega\tau_5}\frac{d\mu}{dt}\bs{B}.
\end{align}
As shown in App.~\ref{app:2}, this expression does not receive finite-temperature corrections up to $O((\mu_5/\mu)(\Delta v/V))$ due to cancelations present for the case of different Fermi velocities. These cancellations are however tied to the particular choice of dispersion relation, as evident from the aCME current  \eqref{eq:b0CME} for a chiral pair of Weyl nodes separated in energy by a amount $b_0$ (c.f.~Fig.~\ref{fig:4})  derived in App.~\ref{app:4}, where no such cancellations happen.

%%%%%%%%%%%%%%%%%%%%%%%%%%%%%%%%%%%%%%%%%%%%%%%%%%%%

\begin{figure}
 \begin{center}
 \includegraphics[width=8.5cm]{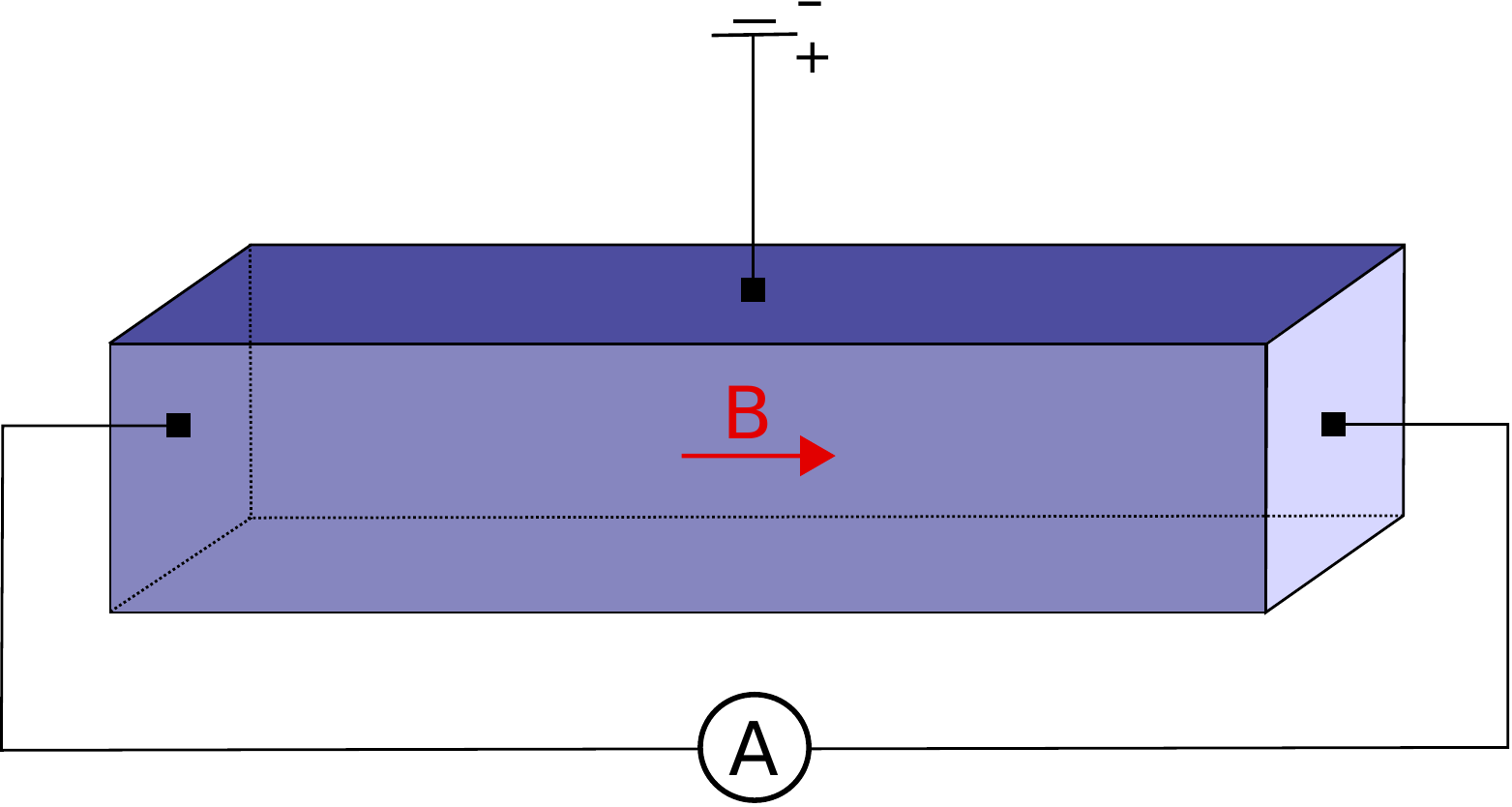}
 \caption{(Color online) Experimental setup to measure the aCME current. The material is put into a homogeneous external magnetic field. The chemical potential is changed by applying a time dependent voltage to the gate, which pumps the system with non chiral electrons. The current is then measured through two gates applied in the direction of the external field. Alternatively, the voltage drop through the sample could be measured instead.}
 \label{fig:exp}
 \end{center}
\end{figure}
\begin{figure}
 \begin{center}
 \includegraphics[width=8.5cm]{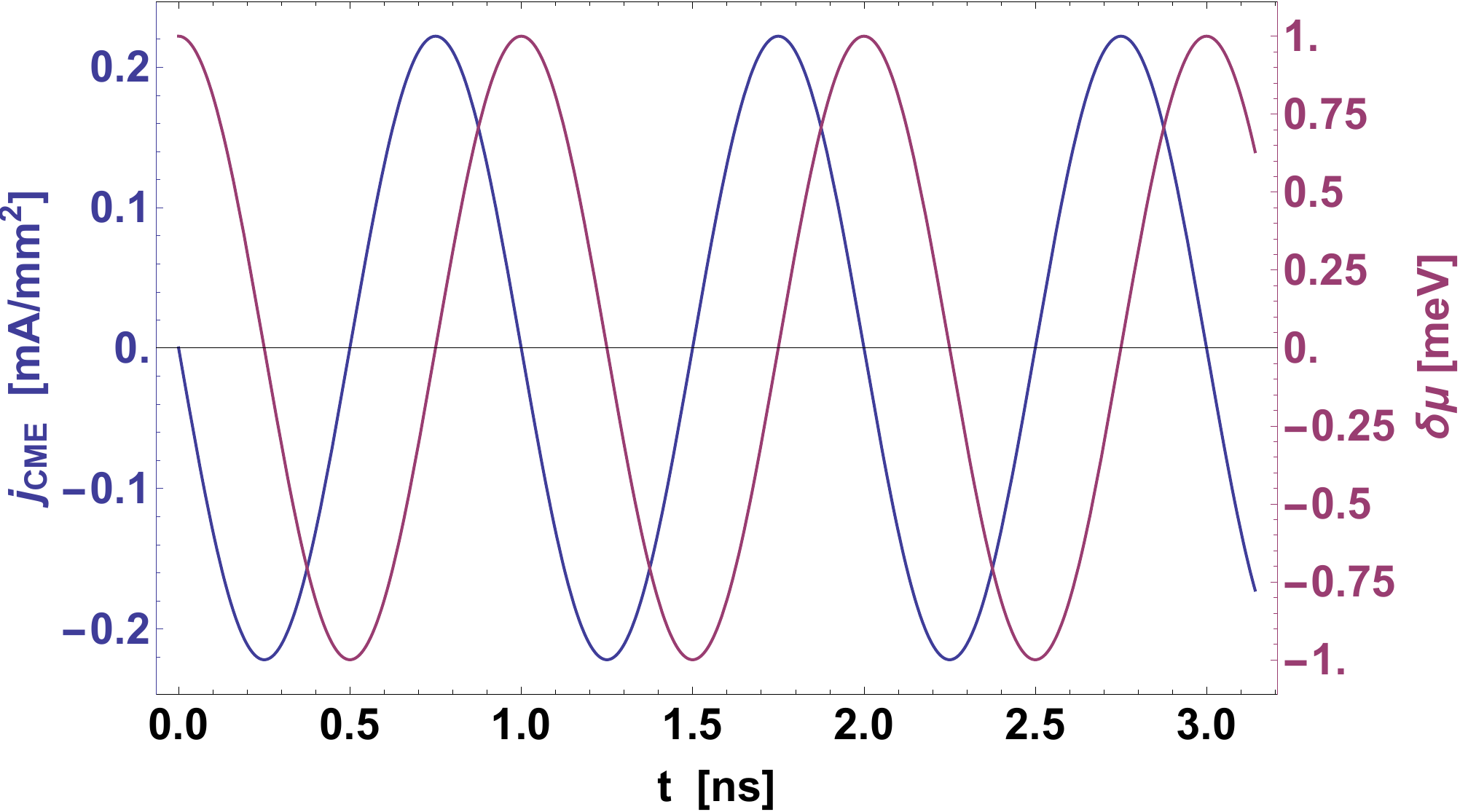}
 \caption{(Color online) Current density response \protect\eqref{eq:CMEA2} for $\Delta = \text{50 meV}$, $\Delta v/V = \text{0.1}$, $|\bs{B}|=\text{1 T}$, and frequencies $\omega = \text{1 GHz}$. Blue curve (the one starting from zero at $t=0$) is the current density response and red curve is time-dependent chemical potential.}
 \label{fig:jcmeplot}
 \end{center}
\end{figure}

\section{\label{sec:5}
Experimental Setup}

 The predictions for the aCME currents \eqref{eq:CME3} and \eqref{eq:b0CME} can be experimentally verified by the simple experimental setup depicted in Fig.~\ref{fig:exp}. The sample, ideally a slab of a single crystal, is connected to a gate providing a reservoir of electrons, and put into a constant external magnetic field. We then expect that once the AC voltage is applied to the gate, the aCME current will be generated parallel to the magnetic field. 
As an electron source, we pump the electric current into the system through spin-polarized contacts. 
There is also a possibility that the chirality imbalance is created by the Fermi arc contribution when the pumped electrons go through the surface of the material even through unpolarized contacts (see \cite{baireuther2016} for related discussion), but we leave the analysis of this mechanism for future work.

 The chemical potential entering \eqref{eq:CME3} and \eqref{eq:b0CME} in this setup is provided through an AC voltage that  pumps electrons into the system at a rate $d\rho/dt$. The chiral relaxation time $\tau_5$ is intrinsic to the system, and has been estimated from the conventional CME in e.g. $\text{ZrTe}_5$ \cite{Li:2014bha} to be parametrically larger than the Ohmic current relaxation rate extracted from the Drude model, $\tau_5 \sim 10\dots 100 \times \tau$ \footnote{This is a necessary condition for the system to be approximately chiral, i.e. a WSM: If the chiral charge would relax faster than the electric current, the chirality of the Weyl fermions would not be approximately preserved on electric transport timescales, and transport would be effectively non-chiral.}. The advantage of this setup is that it does not measure the conventional CME, present only if $\bs{E}$ has a component parallel to $\bs{B}$, but is only sensitive to the new chiral magnetic current  \eqref{eq:CME3}. 

With the result for the aCME current \eqref{eq:CME3} at hand, we can now estimate the current response for aWSMs with approximate rotational invariance but different Fermi velocities on the left and right chiral Weyl cones. Reinstating $\hbar$ and $c$, \eqref{eq:CME3} becomes
\begin{align}
 \label{eq:CME_T}
 \bs{j}_\mathrm{CME} 
 &= \frac{3e^2}{4\pi^2\hbar^2c}\frac{\tau_5}{1+i\omega\tau_5}\frac{\Delta v}{V}\frac{d\mu(t)}{dt}\bs{B}
 \nonumber\\
 & \simeq\frac{3e^2}{4\pi^2\hbar^2c}\tau_5\frac{\Delta v}{V}\frac{d\mu(t)}{dt}\bs{B}. 
\end{align}
Here we have employed the approximation $\omega\tau_5\ll1$, which is the typical regime in the experimental setup we propose here. Since the typical chiral relaxation rates are in the Terahertz regime, $\tau_5^{-1} \sim \frac{O(10\text{meV})}{\hbar} \sim \text{THz}$ \cite{Li:2014bha}, but the technically reachable driving frequencies are $O(\text{GHz})$, $\omega\tau_5 \sim 10^{-3}\ll1$ is well satisfied. 
%The CME current at finite temperature takes the same form as discussed in App.~\ref{app:2}.
%

We now consider AC voltages that induce a time dependent chemical potential given by $\mu(t)=\bar{\mu}+\delta\mu(t)$ with $\delta\mu\ll\bar{\mu}$. This restriction of course limits the amplitude of the driving voltages and hence the aCME current output, but not severely: typical values for the chemical potential (see e.g. the value for $\text{ZrTe}_5$ derived in \cite{Li:2014bha}) are expected to be $O(10\, \text{meV})$, which would allow driving voltages of $0.1\dots 1\, \text{meV}$. We can then use \eqref{eq:rho_mu} to find an approximate linear relationship between the time-dependent part of the chemical potential and charge density, 
\begin{align}
&\bar{\rho}+\delta\rho(t) \simeq \frac{\bar{\mu}^3+E_0^3}{6\pi^2}\left(\frac{1}{v_{L}^3}+\frac{1}{v_{R}^3}\right)
\nonumber\\
& \quad\quad\quad\quad 
+\frac{\bar{\mu}^2}{2\pi^2}\left(\frac{1}{v_{L}^3}+\frac{1}{v_{R}^3}\right)\delta\mu(t).
\end{align}
\begin{figure}
 \begin{center}
 \includegraphics[width=8.5cm]{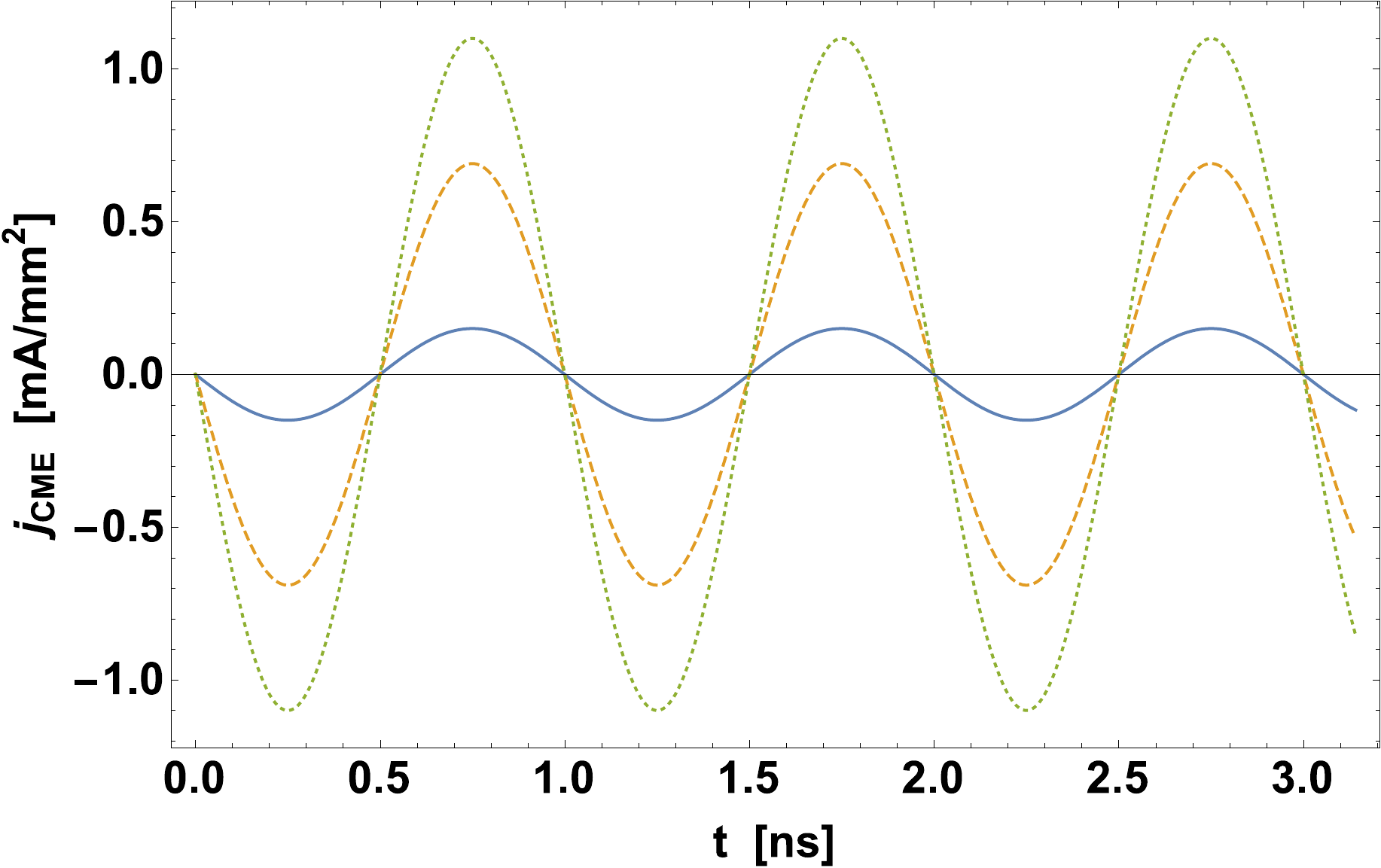}
 \caption{(Color online) Current density response \protect\eqref{eq:b0CME} for $\Delta = \text{50 meV}$, $\nu = \text{1 GHz}$, $|\bs{B}|=\text{1 T}$, $T= \text{20 K}$, $\bar{\mu} = \text{10 meV}$, and $|\delta\mu| = \text{1 meV}$, and each curve corresponds to $b_0 = \text{1 meV}$ (blue solid), $\text{5 meV}$ (yellow dashed) and $\text{10 meV}$ (green dotted).}
 \label{fig:5}
 \end{center}
\end{figure}
\begin{figure}
 \begin{center}
 \includegraphics[width=8.5cm]{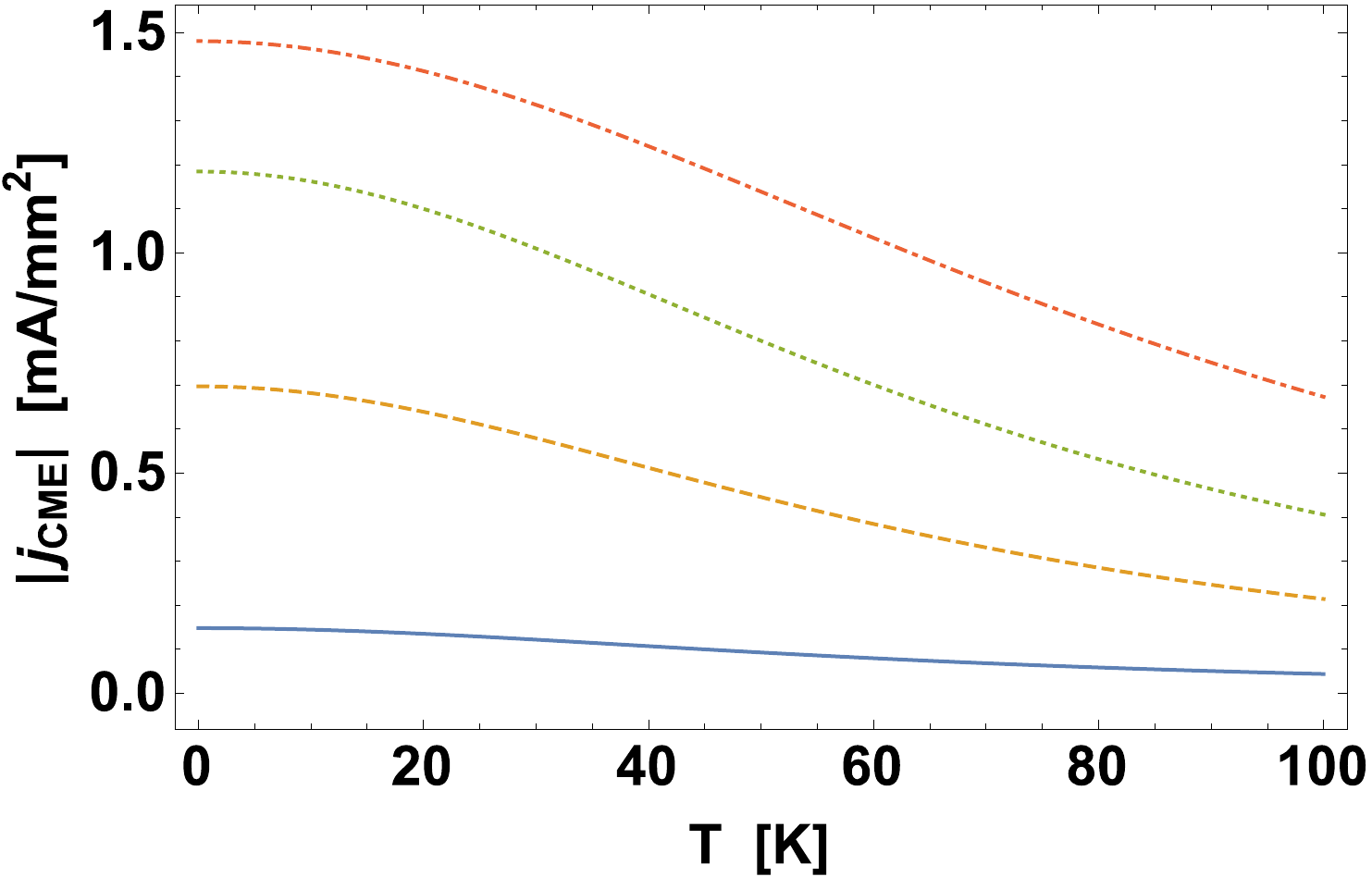}
 \caption{(Color online) Temperature dependence of the amplitudes of current density response \protect\eqref{eq:b0CME} for $\Delta = \text{50 meV}$, $\nu = \text{1 GHz}$, $|\bs{B}|=\text{1 T}$, $\bar{\mu} = \text{10 meV}$, and $|\delta\mu| = \text{1 meV}$, and each curve corresponds to $b_0 = \text{1 meV}$ (blue solid), $\text{5 meV}$ (yellow dashed), $\text{10 meV}$ (green dotted)  and $\text{20 meV}$ (red dotdashed).}
 \label{fig:6}
 \end{center}
\end{figure}
Assuming a harmonically varying chemical potential $\delta\mu(t)\propto\delta\rho(t)\propto e^{i\omega t}$, using $\omega \tau_5 \ll 1$ and taking the real part of \eqref{eq:CME3}, we arrive at (c.f. App.~\ref{app:2} and App.~\ref{app:3} for the derivations of \eqref{eq:CME4} and \eqref{eq:aCMEPara1}) 
\begin{align}
 \label{eq:CME4}
 {\bs{j}_\mathrm{CME}} \,\,\,\,\,\,\,\,
 &= - \frac{3e^2}{4\pi^2\hbar^2c}\frac{\Delta v}{V}
 \left(\omega\tau_5\right) \sin(\omega t)|\delta\mu|\bs{B},\\\nonumber
 \frac{\bs{j}_\mathrm{CME}}{[\text{mA}/\text{mm}^2]} &= -(1.1\times 10^2) 
 \frac{\Delta v}{V}\frac{2\pi\nu}{\mathrm{[GHz]}}\frac{|\delta\mu|}{\mathrm{[meV]}}\left(\frac{\Delta}{\mathrm{[meV]}}\right)^{-1} \\\label{eq:aCMEPara1}
& \quad \times  \frac{\bs{B}}{\mathrm{[T]}}\times  \sin\left( \frac{2\pi\nu}{\mathrm{[GHz]}} \frac{t}{\mathrm{[ns]}} \right).
\end{align}
In the last line, we have converted the chiral relaxation time into an energy scale $\Delta$ via $\tau_5=\hbar/\Delta$ and worked with frequency $\nu$ instead of angular frequency $\omega=2\pi\nu$. 

We can now estimate the magnitude of the asymmetric CME current from \eqref{eq:aCMEPara1}. Using as input parameters $|\delta\mu|=1\,\mathrm{meV}$, $\nu=1\,\mathrm{GHz}$, $\Delta=50\,\mathrm{meV}$, $\Delta v/V=0.1$, and $B=1\,\mathrm{T}$, we find a current density of $|j_{\mathrm{aCME}}| \simeq 20\,\mathrm{mA/cm^2}$.  Note that the parameters in \eqref{eq:CME_T} have to be given in the Gaussian (CGS) units, which are summarized in App.~\ref{app:3}. In App.~\ref{app:3} we also estimate the ordinary CME current with the same input parameters and realistic electric field strengths of $0.1\text{mV}/\text{mm}$ at which the experiments in \cite{Li:2014bha} were performed. The resulting CME current density $|j_{\mathrm{CME}}| = 40\,\mathrm{mA/cm^2}$ is comparable to the current density of our new aCME, which hence should be measurable in laboratory experiments. 

The expected current density response for these input parameters is plotted in Fig.~\ref{fig:jcmeplot} for frequencies $\nu = \mathrm{1\,GHz}\,,\mathrm{2\,GHz}\,,\mathrm{3\,GHz}$, respectively, corresponding to the highest technically attainable frequencies. Since for typical chiral relaxation times $\tau_5 \sim \hbar/(1\dots 10\,\mathrm{meV}) \sim 1\dots 10\ \mathrm{THz}$ this is clearly in the regime $2\pi\nu \tau_5 \ll 1$, we find little additional dependence of the aCME signal on $2\pi\nu \tau_5$ besides the scaling of the amplitude of the signal with $2\pi\nu \tau_5$ as predicted by \eqref{eq:aCMEPara1}: In the regime $2\pi\nu \tau_5 \ll 1$, the novel CME response is predicted to be a sinusoidal signal with the same frequency as the input modulation of the gate voltage, phase shifted by $-\pi$.

In App.~\ref{app:3} we also estimated the aCME signal for the aWSM with left and right Weyl cones separated in energy by an amount $b_0$ (c.f.~Fig.~\ref{fig:4} and Eq.~\eqref{eq:b0CME}). The resulting signal
\begin{align}\label{eq:b0CMEParam}
 &\frac{\bs{j}_\mathrm{CME}}{\mathrm{[mA/mm^2]}}
 = -(7.4\times 10)
 \nonumber\\
 &\times\left(\frac{\bar{\mu}}{b_0}+0.024\times\left(\frac{T}{\mathrm[K]}\right)^2\left(\frac{b_0\bar{\mu}}{[\mathrm{meV}^2]}\right)^{-1}+\frac{b_0}{4\bar{\mu}}\right)^{-1}
 \nonumber\\
 &\times\frac{\omega}{\mathrm{[GHz]}}\frac{|\delta\mu|}{\mathrm{[meV]}}\left(\frac{\Delta}{\mathrm{[meV]}}\right)^{-1}
 \frac{\bs{B}}{\mathrm{[T]}}
 \sin\left( \frac{\omega}{\mathrm{[GHz]}} \frac{t}{\mathrm{[ns]}} \right)
\end{align}
is plotted in Fig.~\ref{fig:5} for comparable input parameters $\Delta = \text{50 meV}$, $\nu = \text{1 GHz}$, $|\bs{B}|=\text{1 T}$, $T= \text{20 K}$, $\bar{\mu} = \text{10 meV}$, and $|\delta\mu| = \text{1 meV}$. The resulting aCME current density is generally of the same order of magnitude as for the case of unequal Fermi velocities, and for adiabatic pumping by a cosine input signal, the output is again shifted by a phase $-\pi$, i.e. is a negative sine with the same frequency as the input signal. The temperature dependence of $|\bs{j}_\mathrm{CME}|$, which is plotted in Fig.~\ref{fig:6} for different values of $b_0$, is found to be small for small $b_0$ but significantly enhanced for larger values of $b_0$.

%%%%%%%%%%%%%%%%%%%%%%%%%%%%%%%%%%%%%%%%%%%%%%%%%%%%
\section{\label{sec:6}
Conclusions}

The chiral magnetic effect \eqref{eq1} is a non dissipative transport phenomenon induced by the chiral anomaly in chirally imbalanced systems, i.e. in non equilibrium states with finite chiral chemical potential $\mu_5$. In the conventional realization of CME in condensed matter experiments, the chirality imbalance is generated via the chiral anomaly itself by applying external electromagnetic fields with 
non vanishing $\bs{E}\cdot\bs{B}$. In this paper we presented a novel chiral magnetic effect which does not rely on an external source of chirality imbalance. Instead, the source of chirality imbalance is built into the band structure of certain materials, asymmetric Weyl semimetals ({aWSMs}), that break parity by means of asymmetric, i.e. non-identical, dispersion relations of their left and right chiral Weyl cones. The chirality imbalance is then generated upon pumping the system with a time-dependent non chiral current, which changes the chemical potential in  the left and right chiral Weyl cones at a different rate due to their different capacities. 

 The aCME current induced by this chiral chemical potential will flow along the applied magnetic field according to \eqref{eq1}, even if no electric field at all is applied to the system, or if the electric field is perpendicular to the magnetic field. 
 In Sec.~\ref{sec:5}, we  presented a detailed experimental setup to measure the new aCME in asymmetric WSM,  and estimated the induced currents quantitatively (see App.~\ref{app:3} for details). For realistic input parameters for the chemical potential, temperature and the chiral relaxation rate, we found an aCME current of the same order of magnitude, and possibly even stronger, as the conventional CME current. For our input parameters we assumed that the values inferred for the DSM ZrTe${}_5$ \cite{Li:2014bha} are typical also for aWSMs. 
 
We have to admit that our estimate of the value of the chiral chemical potential (and thus of the magnitude of the chiral magnetic current) crucially depends on the assumptions about the physics of electron pumping through the contacts. {However, once created, the population asymmetry will persist in a surface layer of thickness determined by the chirality 
flipping length ($v_F \tau_5$ in the ballistic regime, or $\sqrt{D \tau_5}$ in the diffusion regime). Experimentally, in CdAs$_3 $ \cite{Zhang:2017dd}
it is known to be $2$-$3 \mu$m. We hence expect our effect to persist for films up to this thickness, which otherwise are known to  behave as bulk Weyl semimetals.} The existence of the current parallel to magnetic field and orthogonal to the applied electric field, and the shift of $\pi/2$ in phase between the chiral magnetic current and the driving voltage then make it possible to observe the effect in experiment.
 %%%%%%%%%
\vskip0.3cm

Apart from the AC voltage considered above, one could also drive the system by  laser. In this case the laser light will induce oscillations of charge density which can have much higher frequencies than the AC source, e.g. in the terahertz frequency range. This would yield a much stronger aCME current. {The effect is also calculable by computing the bulk AC conductivity tensor in linear response in the limit of large chirality flipping length. Since it has been shown in several examples, including the DC and AC chiral magnetic conductivity \cite{Fukushima:2008xe,PhysRevD.80.034028,Kharzeev:2016sut}, that chiral kinetic theory agrees with linear response theory, we expect our result to be reproduced in this way.}
\vskip0.3cm

Our effect is predicted to exist in materials in which the left- and right-handed Weyl quasiparticles have different dispersion relations. {Furthermore, it requires the absence of any mirror planes (reflection symmetries), as reflections interchange the chiralities of the Weyl nodes and hence of any relative asymmetry in their dispersion relations, such that the predicted effect would cancel between the asymmetric Weyl pair and its mirror image. These requirements are not very restrictive, and many such materials should exist. In terms of the Schoenflies classification of point groups, the allowed classes are $C_n$ (but not $C_{nh}$ and $C_{nv}$), $D_n$ (but not $D_{nd}$ and $D_{nh}$), $T$ (but not $T_d$ and $T_h$), $O$ (but not $O_h$) and $I$ (but not $I_h$).}
Recently, such Weyl semimetals with broken reflection and inversion symmetries were predicted theoretically, with 
${\rm SrSi_2}$ {(space group classification $P4_3 32$)} being an example \cite{huang2016new} {in the Schoenflies class $O$}. We urge the experimental study of aCME in this and other asymmetric Weyl semimetals, as this would greatly improve the understanding of the role that chiral anomaly plays in transport phenomena.

%%%%%%%%%%%%%%%%%%%%%%%%%%%%%%%%%%%%%%%%%%%%%%%%%%%%
\acknowledgement
We would like to thank Karl Landsteiner, Qiang Li, Shinsei Ryu, Maria Vozmediano and Yi Yin for useful discussions. The work of DK was supported in part by the U.S.
Department of Energy under Contracts No. DE-FG- 88ER40388 and DE-AC02-98CH10886, and by the LDRD 16-004 at Brookhaven National Laboratory. 
Y.K. is supported by the Grants-in-Aid for JSPS fellows (Grant No.15J01626). The work of R.M. was supported in part by the U.S. Department of Energy 
under Contract No. DE-FG-88ER40388, as well as by the Alexander-von-Humboldt Foundation
through a Feodor Lynen postdoctoral fellowship.
\newline

{Contribution statement: The three authors, D.K., Y.K. and R.M., contributed equally to this work.}

%%%%%%%%%%%%%%%%%%%%%%%%%%%%%%%%%%%%%%%%%%%%%%%%%%%%

%\newpage

\appendix

%%%%%%%%%%%%%%%%%%%%%%%%%%%%%%%%%%%%%%%%%%%%%%%%%%%%
\section{\label{app:1}
Computation of index for aWSMs with different Fermi velocities}

In this appendix, we present the detailed computation of the index for the left chiral fermions \eqref{eq:left_CS}.
To this end, we introduce a cutoff $\Lambda$ which regularizes the states with large Dirac eigenvalues,
\begin{align}
 \label{eq:index_comp1}
 &\sum_n\phi^*_n(x)\gamma_5\phi_n(x)
 \nonumber\\
 &=\lim_{\Lambda\to\infty}\left(\sum_n\phi^*_n(x)\gamma_5e^{-(\lambda_n/\Lambda)^2}\phi_n(x)\right)
 \nonumber\\
 &=\lim_{\Lambda\to\infty}\mathrm{tr}\left[\gamma_5e^{-(\slashed{D}_L/\Lambda)^2}\sum_n\phi_n(x)\phi^*_n(x)\right]
 \nonumber\\
 &=\lim_{\Lambda\to\infty}\int\frac{d^4k}{(2\pi)^4}\mathrm{tr}\left[\gamma_5 e^{-ikx}e^{-(\slashed{D}_L/\Lambda)^2}e^{ikx}\right],
\end{align}
where we have changed the basis to plane waves by,
\begin{align}
 \sum_n\phi_n(x)\phi^*_n(y)=\bs{1}_{4\times4}\delta(x-y)=\bs{1}_{4\times4}\int\frac{d^4k}{(2\pi)^4}e^{-ik(x-y)}.
\end{align}
$\bs{1}_{4\times4}$ is the identity matrix in the spinor space.
Notice that $\slashed{D}_L$, defined in \eqref{eq:DL_DR}, explicitly depends on the Fermi velocity and $\slashed{D}_L^2$ is expanded as
\begin{align}
 \slashed{D}_L^2 &=-D_0D_0-v_L^2 D_iD_i -b_0b_0-v_L^2 b_ib_i 
 \nonumber\\
 &+ \frac{ie}{4}(2v_L[\gamma^0,\gamma^i]F_{0i} + v_L^2[\gamma^i,\gamma^j]F_{ij})
 \nonumber\\
 &+ i\gamma_5(v_L[\gamma^0,\gamma^i](b_0D_i-b_iD_0) + v_L^2[\gamma^i,\gamma^j]b_iD_j).
\end{align}
Therefore, \eqref{eq:index_comp1} can be calculated as
\begin{align}
 \label{eq:index_app}
 &\sum_n\phi^*_n(x)\gamma_5\phi_n(x)
 \nonumber\\
 &=\lim_{\Lambda\to\infty}\int\frac{d^4k}{(2\pi)^4}
 \nonumber\\
 &\times\mathrm{tr}\Bigg[\gamma_5
 \exp\Bigg(\frac{(ik_0+D_0)^2+v_L^2(ik_i+D_i)^2}{\Lambda^2}
 +\frac{b_0^2+v_L^2b_i^2}{\Lambda^2}
 \nonumber\\
 &-\frac{ie}{4\Lambda^2}(2v_L[\gamma^0,\gamma^i]F_{0i} + v_L^2[\gamma^i,\gamma^j]F_{ij})
 \nonumber\\
 &-\frac{i}{\Lambda^2}\gamma_5(v_L[\gamma^0,\gamma^i](b_0D_i-b_iD_0) + v_L^2[\gamma^i,\gamma^j]b_iD_j)\Bigg)\Bigg]
 \nonumber\\
 &=\frac{1}{16\pi^2v_L^3}
 \nonumber\\
 &\times \mathrm{tr}\frac{1}{2}\Bigg[\gamma_5
 \Bigg(\frac{ie}{4}(2v_L[\gamma^0,\gamma^i]F_{0i} + v_L^2[\gamma^i,\gamma^j]F_{ij})\Bigg)^2\Bigg]
 \nonumber\\
 &=\frac{e^2}{32\pi^2}\epsilon^{\mu\nu\alpha\beta}F_{\mu\nu}F_{\alpha\beta}.
\end{align}
In the final equality, we have used the identity
\begin{align}
 \mathrm{tr}[\gamma^5\gamma^\mu\gamma^\nu\gamma^\alpha,\gamma^\beta]=-4\epsilon^{\mu\nu\alpha\beta},
\end{align}
in Euclidean signature.
Note that $v_L$ drops out due to picking up the cross term in the final equality of \eqref{eq:index_app}.

%%%%%%%%%%%%%%%%%%%%%%%%%%%%%%%%%%%%%%%%%%%%%%%%%%%%

%%%%%%%%%%%%%%%%%%%%%%%%%%%%%%%%%%%%%%%%%%%%%%%%%%%%

\section{\label{app:2}
Finite temperature expressions for aCME with different Fermi velocities}

In this appendix, we present the finite temperature expression of the aCME current.
The left Weyl node contribution to the current is given by
\begin{align}
 &\bs{j}_{\mathrm{CME},L} 
 = e^2\bs{B}\int_{\bs{p}} f_{\bs{p}} \left(\bs{\Omega}\cdot\frac{\partial \epsilon_{\bs{p}}}{\partial\bs{p}}\right)
 \nonumber\\
 &= \frac{e^2\bs{B}}{4\pi^2}T\ln(1+e^{(\mu_L+E_0)/T})
 \nonumber\\
 &= \frac{e^2\bs{B}}{4\pi^2}(\mu_L+E_0)(1+ O(e^{-(\mu_L+E_0)/T})),
\end{align}
where the distribution function is $f_{\bs{p}}=1/(1+\mathrm{e}^x)$ with $x=(\epsilon_{\bs{p}}-\mu_L)/T$.
Notice that the nontrivial temperature dependence is exponentially suppressed is suppressed at low temperatures due to the Fermi sea cutoff $E_0$ \cite{Basar:2013iaa}.
Hence the net CME current is expressed as
\begin{align}
 &\bs{j}_{\mathrm{CME}} = \frac{e^2\bs{B}}{2\pi^2}\mu_5.
\end{align}
Therefore, the expression for CME current is unchanged up to the exponentially small corrections at low temperatures.

The left and right chiral fermion densities receive a finite temperature correction
\begin{align}
 \rho_{L/R} = \frac{\mu_{L/R}^3+\pi^2 T^2\mu_{L/R}+E_0^3}{6\pi^2v_{L/R}^3}(1+ O(e^{-(\mu_{L/R}+E_0)/T})).
\end{align}
At finite temperature, Eqs.~\eqref{eq:rhomu1} and \eqref{eq:rhomu2} become
\begin{align}
 v_L^3\delta\rho_L-v_R^3\delta\rho_R
 &=\frac{\tau_5}{1+i\omega\tau_5}\frac{d\rho(t)}{dt}\frac{v_L^3-v_R^3}{2},
 \nonumber\\
 &= \left(\frac{\mu^2}{\pi^2}+\frac{T^2}{3}\right)\mu_5.
\end{align}
The finite temperature version of \eqref{eq:mu5i} can be read off,
\begin{align}
 \label{eq:B5}
 \mu_5= \frac{\pi^2\hbar^3(v_L^3-v_R^3)\tau_5}{2(\mu^2+\pi^2T^2/3)}\frac{\tau_5}{1+i\omega\tau_5}\frac{d\rho(t)}{dt}.
\end{align}
The relation between the particle number density and chemical potential is
\begin{align}
&\rho(t) 
\nonumber\\
&= \frac{1}{6\pi^2\hbar^3}\left(\frac{\mu_L^3+\pi^2T^2\mu_L+E_0^3}{v_{L}^3}+\frac{\mu_R^3+\pi^2T^2\mu_R+E_0^3}{v_{R}^3}\right)
\nonumber\\
&= \frac{\mu^3+\pi^2T^2\mu+E_0^3}{6\pi^2\hbar^3}\left(\frac{1}{v_{L}^3}+\frac{1}{v_{R}^3}\right)\left(1+O\left(\frac{\Delta v}{V}\frac{\mu_5}{\mu}\right)\right),
\end{align}
which leads to
\begin{align}
 \label{eq:B7}
&\frac{d\rho(t)}{dt} 
= \frac{3\mu^2+\pi^2T^2}{6\pi^2\hbar^3}\left(\frac{1}{v_{L}^3}+\frac{1}{v_{R}^3}\right)\frac{d\mu}{dt}\left(1+O\left(\frac{\Delta v}{V}\frac{\mu_5}{\mu}\right)\right).
\end{align}
Therefore, the chiral chemical potential is
\begin{align}
 \mu_5= \frac{(v_L^6-v_R^6)}{4v_L^3v_R^3}\frac{d\mu(t)}{dt}\tau_5
 \simeq \frac{3}{2}\frac{\Delta v}{V}\frac{\tau_5}{1+i\omega\tau_5}\frac{d\mu(t)}{dt}.
\end{align}
Finally, the aCME current is given by
\begin{align}
 \label{eq:CMEA}
 \bs{j}_\mathrm{CME} 
 &= \frac{3e^2}{4\pi^2\hbar^2c}\frac{\Delta v}{V}\frac{\tau_5}{1+i\omega\tau_5}\frac{d\mu}{dt}\bs{B}.
\end{align}
By using $\mu(t)=\bar{\mu}+\delta\mu(t)$ with $\delta\mu(t)\propto e^{i\omega t}$ and taking the real part of \eqref{eq:CMEA},
\begin{align}
 \label{eq:CMEA2}
 \bs{j}_\mathrm{CME} 
 &= \frac{3e^2}{4\pi^2\hbar^2c}\frac{\Delta v}{V}
 \frac{(\omega\tau_5)^2\cos\omega t-\omega\tau_5\sin(\omega t)}{1+(\omega\tau_5)^2}|\delta\mu|\bs{B}.
\end{align}
Note that to $O((\Delta v/V)(\mu_5/\mu))$, the temperature dependence completely cancelled between \eqref{eq:B5} and \eqref{eq:B7}. This is a special feature of the chosen dispersion and approximation, which e.g. does not happen in the energy separated aWSM of App.\ref{app:3}.

%%%%%%%%%%%%%%%%%%%%%%%%%%%%%%%%%%%%%%%%%%%%%%%%%%%%

%%%%%%%%%%%%%%%%%%%%%%%%%%%%%%%%%%%%%%%%%%%%%%%%%%%%

\section{\label{app:4}
aCME with energy separation between Weyl nodes}

The aCME is also realized in materials with finite energy separation between the left and right Weyl nodes, denoted by $b_0$, as depicted in Fig.\ref{fig:4}.
In this appendix, we present the aCME expression with finite $b_0$ instead of different right and left Fermi velocities.
The particle number densities of the left and right Weyl fermions are given by
\begin{align}\label{eq:D1}
 &\rho_{L/R} 
 \nonumber\\
 &= \frac{(\mu_{L/R}\mp b_0/2)^3+\pi^2 T^2(\mu_{L/R}\mp b_0/2)+(E_0\pm b_0/2)^3}{6\pi^2v_{F}^3}.
\end{align}

Correspondingly, $\rho_{L,R}^{\mathrm{CB}}$ and $\rho_5^{\mathrm{CB}}$ are given by
\begin{align}
 &\rho_{L/R}^{\mathrm{CB}}
 \nonumber\\
 &= \frac{(\mu_{\mathrm{CB}}\mp b_0/2)^3+\pi^2 T^2(\mu_{\mathrm{CB}}\mp b_0/2)+(E_0\pm b_0/2)^3}{6\pi^2v_{F}^3},
 \\
 &\rho_{5}^{\mathrm{CB}}
 = -\frac{b_0(\mu_{\mathrm{CB}}^2+\pi^2 T^2/3-E_0^2)}{2\pi^2v_{F}^3}.
\end{align}
Hence we obtain
\begin{align}
 \frac{d\delta\rho_5}{dt}
 &= \frac{b_0}{2\pi^2v_{F}^3}\frac{d\mu_{\mathrm{CB}}^2}{dt}
 -\frac{\delta\rho_5}{\tau_5}
 \nonumber\\
 &\simeq \frac{b_0\bar{\mu}}{\pi^2v_{F}^3}\frac{d\delta\mu}{dt}
 -\frac{\delta\rho_5}{\tau_5},
\end{align}
where we have used $\mu_{\mathrm{CB}}=\mu(1+O(\mu_5/\mu))$ and 
$\mu=\bar{\mu}+\delta\mu$ with $|\delta\mu|\ll\bar{\mu}$.
By assuming $\delta\mu\propto e^{i\omega t}$, we obtain the solution,
\begin{align}
 \delta\rho_5
 &= \frac{b_0\bar{\mu}}{\pi^2v_{F}^3}\frac{\tau_5}{1+i\omega\tau_5}\frac{d\mu}{dt},
\end{align}
In terms of $\mu_5$, $\delta\rho_5$ can be expressed as
\begin{align}
 \delta\rho_5=\frac{\mu^2+\pi^2T^2/3+b_0^2/4}{\pi^2v_F^3}\mu_5
\end{align}
Therefore, the chiral chemical potential becomes
\begin{align}
 \mu_5= \frac{b_0\bar{\mu}}{\mu^2+\pi^2T^2/3+b_0^2/4}\frac{\tau_5}{1+i\omega\tau_5}\frac{d\mu}{dt},
\end{align}
$\mu_5\ll\mu$ is readily satisfied for $\omega\tau_5\ll1$.
Then, the induced CME current is expressed as
\begin{align}
 \label{eq:b0CME}
 \bs{j}_{\mathrm{CME}}&= \frac{e^2}{2\pi^2\hbar^2c}\frac{b_0\bar{\mu}}{\bar{\mu}^2+\pi^2T^2/3+b_0^2/4}
 \nonumber\\
 &\times\frac{(\omega\tau_5)^2\cos\omega t-\omega\tau_5\sin(\omega t)}{1+(\omega\tau_5)^2}|\delta\mu|\bs{B}
 \nonumber\\
 &\simeq -\frac{e^2}{2\pi^2\hbar^2c}\frac{b_0\bar{\mu}}{\bar{\mu}^2+\pi^2T^2/3+b_0^2/4}
 |\delta\mu|\omega\tau_5\sin(\omega t)\bs{B}.
\end{align}

%%%%%%%%%%%%%%%%%%%%%%%%%%%%%%%%%%%%%%%%%%%%%%%%%%%%

%%%%%%%%%%%%%%%%%%%%%%%%%%%%%%%%%%%%%%%%%%%%%%%%%%%%

\section{\label{app:3}
Parameterization for CME current}

The parameters in Eqs.~\eqref{eq:CMEA} and \eqref{eq:b0CME}
are given in Gaussian (CGS) units.
The parameterizations given in Sec.~\ref{sec:5} are expressed in Gaussian units as follows:
\begin{align}
 |\delta\mu|&=1\,\mathrm{meV} = 1.6\times10^{-15}\,\mathrm{erg},
 \\
 \Delta&=50\,\mathrm{meV} = 8\times10^{-14}\,\mathrm{erg},
 \\
 B&=1\,\mathrm{T} = 1\times 10^4\, \mathrm{Gs},
 \\
 T&=20\/\mathrm{K} = 1.4\times 10^{-16} \, \mathrm{erg}.
\end{align}
Furthermore $e=4.8\times10^{-10}\,\mathrm{Fr}$, $\hbar=1.1\times10^{-27}\,\mathrm{erg\,s}$, and $c=3\times10^{10}\,\mathrm{cm/s}$.

The result \eqref{eq:CMEA} expressed in SI units reads,
\begin{align}
 &\frac{\bs{j}_\mathrm{CME}}{\mathrm{[A/m^2]}}(3\times10^5) 
 \nonumber\\
 &= -\frac{3(4.8\times10^{-10})^2}{(6.6\times10^{-27})^2(3\times10^{10})}
 \nonumber\\
 &\times\frac{\Delta v}{V}\frac{\frac{\omega}{\mathrm{[GHz]}}(6.6\times10^{-18})\frac{|\delta\mu|}{\mathrm{[meV]}}(1.6\times10^{-15})}{\frac{\Delta}{\mathrm{[meV]}}(1.6\times10^{-15})}
 \nonumber\\
 &\times\frac{\bs{B}}{\mathrm{[T]}}(1\times10^4)
 \sin\left( \frac{\omega}{\mathrm{[GHz]}} \frac{t}{\mathrm{[ns]}} \right),
\end{align}
which can be simplified to
\begin{align}
 &\frac{\bs{j}_\mathrm{CME}}{\mathrm{[mA/mm^2]}}
 \nonumber\\
 &= -(1.1\times10^2)
 \frac{\Delta v}{V}\frac{\omega}{\mathrm{[GHz]}}\frac{|\delta\mu|}{\mathrm{[meV]}}\left(\frac{\Delta}{\mathrm{[meV]}}\right)^{-1}
 \frac{\bs{B}}{\mathrm{[T]}}
 \nonumber\\
 &\times\sin\left( \frac{\omega}{\mathrm{[GHz]}} \frac{t}{\mathrm{[ns]}} \right).
\end{align}
The amplitude of the induced current is calculated to be
\begin{align}
 &7.0\times10^{7}\,\mathrm{Fr/(s\,cm^2)} = 2.2\times 10^{2}\,\mathrm{A/m^2}
 \nonumber\\
 &=2.2\times{10^{-1}}\,\mathrm{mA/mm^2}.
\end{align}

On the other hand, the result \eqref{eq:b0CME} in SI units reads,
\begin{align}
 &\frac{\bs{j}_\mathrm{CME}}{\mathrm{[A/m^2]}}(3\times10^5) 
 = -\frac{2(4.8\times10^{-10})^2}{(6.6\times10^{-27})^2(3\times10^{10})}
 \nonumber\\
 &\times\frac{1}{\frac{\bar{\mu}}{\mathrm{[meV]}}/\frac{b_0}{\mathrm{[meV]}}+(0.086\pi\times\frac{T}{\mathrm[K]})^2/\frac{3b_0\bar{\mu}}{\mathrm{[meV]}^2}+\frac{b_0}{\mathrm{[meV]}}/\frac{4\bar{\mu}}{\mathrm{[meV]}}}
 \nonumber\\
 &\times\frac{\frac{\omega}{\mathrm{[GHz]}}(6.6\times10^{-18})\frac{|\delta\mu|}{\mathrm{[meV]}}(1.6\times10^{-15})}{\frac{\Delta}{\mathrm{[meV]}}(1.6\times10^{-15})}
 \frac{\bs{B}}{\mathrm{[T]}}(1\times10^4)
 \nonumber\\
 &\times\sin\left( \frac{\omega}{\mathrm{[GHz]}} \frac{t}{\mathrm{[ns]}} \right),
\end{align}
which can be simplified to
\begin{align}
 &\frac{\bs{j}_\mathrm{CME}}{\mathrm{[mA/mm^2]}}
 = -(7.4\times 10)
 \nonumber\\
 &\times\left(\frac{\bar{\mu}}{b_0}+0.024\times\left(\frac{T}{\mathrm[K]}\right)^2\left(\frac{b_0\bar{\mu}}{[\mathrm{meV}^2]}\right)^{-1}+\frac{b_0}{4\bar{\mu}}\right)^{-1}
 \nonumber\\
 &\times\frac{\omega}{\mathrm{[GHz]}}\frac{|\delta\mu|}{\mathrm{[meV]}}\left(\frac{\Delta}{\mathrm{[meV]}}\right)^{-1}
 \frac{\bs{B}}{\mathrm{[T]}}
 \sin\left( \frac{\omega}{\mathrm{[GHz]}} \frac{t}{\mathrm{[ns]}} \right).
\end{align}

In addition to the aCME current, we consider the conventional CME current for comparison,
\begin{align}
 \bs{j}_{\mathrm{CME}}=\frac{e^4v_F^3}{8\pi^2\hbar c^2}\frac{\tau_5}{(\mu^2+\pi^2T^2)}(\bs{E}\cdot\bs{B})\bs{B}.
\end{align}
In SI units, this reads
\begin{align}
 &\frac{\bs{j}_{\mathrm{CME}}}{\mathrm{[A/m^2]}}(3\times10^5) 
 \nonumber\\
 &= \frac{(4.8\times10^{-10})^4(3\times10^{10}/300)^3}{8\pi^2(3\times10^{10})^2}
 \frac{1}{\frac{\Delta}{\mathrm{[meV]}}(1.6\times10^{-15})}
 \nonumber\\
 &\times
 \frac{1}{\left(\frac{\mu}{\mathrm{[meV]}}(1.6\times10^{-15})\right)^2+\pi^2\left(\frac{T}{\mathrm{[K]}}(1.4\times10^{-16})\right)^2}
 \nonumber\\
 &\times\frac{E}{\mathrm{[V/m]}}(3\times10^4)^{-1}\left(\frac{B}{\mathrm{[T]}}(1\times10^4)\right)^2
 \nonumber\\
 &\times\sin\left( \frac{\omega}{\mathrm{[GHz]}} \frac{t}{\mathrm{[ns]}} \right),
\end{align}
which is simplified to
\begin{align}
 &\frac{\bs{j}_{\mathrm{CME}}}{\mathrm{[mA/mm^2]}}
 =(2.0\times 10^3)
 \left(\frac{\Delta}{\mathrm{[meV]}}\right)^{-1}
 \nonumber\\
 &\times\left(\left(\frac{\mu}{\mathrm{[meV]}}\right)^2+(7.6\times 10^{-2})\left(\frac{T}{\mathrm{[K]}}\right)^2\right)^{-1}
 \frac{E}{\mathrm{[V/m]}}\left(\frac{B}{\mathrm{[T]}}\right)^2
 \nonumber\\
 &\times\sin\left( \frac{\omega}{\mathrm{[GHz]}} \frac{t}{\mathrm{[ns]}} \right).
\end{align}
For $\Delta=50\mathrm{[meV]}$, $\mu=10\mathrm{[meV]}$, $T=0\mathrm{[K]}$, $E=0.1\mathrm{[mV/mm]}$, and $B=1\mathrm{[T]}$, the amplitude of the current becomes $4.0\times10^{-1}\mathrm{[mA/mm^2]}$. We obtain the amplitude $4.0\times10\mathrm{[mA/mm^2]}$ if we increase the magnetic field to $B=10\mathrm{[T]}$.

%%%%%%%%%%%%%%%%%%%%%%%%%%%%%%%%%%%%%%%%%%%%%%%%%%%%

%%%%%%%%%%%%%%%%%%%%%%%%%%%%%%%%%%%%%%%%%%%%%%%%%%%%
\bibliographystyle{unsrt}
\bibliography{references}% Produces the bibliography via BibTeX.

\end{document}